\documentclass[acmsmall,nonacm]{acmart} % nonacm for preprint
\AtBeginDocument{%
  }

\usepackage{booktabs}
\usepackage{multirow}
\usepackage{xspace}
\usepackage{tcolorbox}
\usepackage{amsfonts} % for \mathbb
\usepackage{colortbl}
\usepackage{subcaption}
\usepackage{wrapfig}

\newcommand{\manualsubcaption}[2]{%
  \refstepcounter{subfigure}%
  \label{#1}%
  \par\vspace{0.25em}%
  {\small\centering(\thesubfigure)~#2\par}%
}

\definecolor{diagramred}{HTML}{EE220C} 
\definecolor{diagramteal}{HTML}{00AB8E}

\usepackage{listings}
\definecolor{dkgreen}{rgb}{0,0.5,0}
\definecolor{lessdkgreen}{rgb}{0,0.6,0}
\definecolor{dkred}{rgb}{0.5,0,0}
\definecolor{gray}{rgb}{0.5,0.5,0.5}
\definecolor{dkblue}{rgb}{0,0,0.7}

\lstdefinestyle{cstyle}{
language=c,
basicstyle=\fontencoding{T1}\fontsize{7}{8}\selectfont\ttfamily\bfseries,
  morekeywords={virtualinvoke},
  keywordstyle=\color{blue},
  ndkeywordstyle=\color{red},
  commentstyle=\color{dkred},
  stringstyle=\color{dkgreen},
  numbers=left,
  breaklines=true,
  numberstyle=\ttfamily\fontsize{6}{7}\color{gray},
  stepnumber=1,
  numbersep=4pt,
  backgroundcolor=\color{white},
  tabsize=4,
  showspaces=false,
  showstringspaces=false,
  xleftmargin=.2in,
  captionpos=b,
  print
}
\lstdefinestyle{cinlinestyle}{
language=c,
basicstyle=\fontencoding{T1}\ttfamily\footnotesize,  % \bfseries
  morekeywords={virtualinvoke},
  keywordstyle=\color{blue},
  ndkeywordstyle=\color{red},
  commentstyle=\color{dkred},
  stringstyle=\color{dkgreen},
  print
}

\newcommand\cinline[1]{{\lstinline[style=cinlinestyle]!#1!}}

\usepackage{fancybox}

\newcommand{\realtype}{\textsc{realtype}\xspace}
\newcommand{\idioms}{\textsc{idioms}\xspace}
\newcommand{\exebench}{\textsc{exebench}\xspace}
\newcommand{\faultless}{\textsc{faultless}\xspace}

\setcopyright{acmlicensed}
\copyrightyear{2026}
\acmYear{2026}
\begin{document}

\title{Faultless: A Program Equivalence Technique for Validating and Evaluating Neural Decompilers}

\author{Luke Dramko}
\correspondingauthor
\email{lukedram@cs.cmu.edu}
\author{Claire Le Goues}
\email{clegoues@cs.cmu.edu}
\affiliation{%
  \institution{Carnegie Mellon University}
  % \city{Pittsburgh}
  \country{USA}
}

\author{Edward Schwartz}
\email{eschwartz@cert.org}
\affiliation{%
  \institution{Carnegie Mellon University Software Engineering Institute}
 % \city{Rocquencourt}
  \country{USA}
}

\renewcommand{\shortauthors}{Dramko et al.}

\begin{abstract}
Neural decompilers are machine learning models which perform the process of decompilation, lifting code from a lower-level language to a higher one.
Neural decompilers offer substantial utility relative to traditional deterministic decompilers because they can probabilistically recover information discarded during lowering, like variable names, types, and control flow structuring.
However, they can also make mistakes, producing code that is not equivalent to the original, making it difficult to trust their output.

In this work, we introduce \faultless, a program equivalence technique for performing translation validation on neural decompilers.
\faultless compares code produced by a deterministic decompiler, which has stronger correctness properties, with that of a neural decompiler.
\faultless is also useful for model evaluation, a highly related task, in which the neural decompilers' prediction is compared with a reference solution.
Neural decompilation introduces significant challenges to the task of program equivalence which existing techniques are not equipped to handle, including limited extrafunctional context and systematic semantic inconsistencies in decompiled code.
\faultless takes a static symbolic execution-based approach with an execution model and memory model designed to handle these challenges.
\end{abstract}




\maketitle

\section{Introduction}

Decompilers are tools which raise the abstraction level of a program from a lower level (like executable code) to a higher level (like C).
Reverse engineers use decompilers for important security tasks like malware analysis and vulnerability research.
Unfortunately, decompilation often fails to exactly reconstruct the original source code because compilation is lossy: variable names, some type information, structured control flow, comments, and other features of source code are discarded during compilation and not preserved in the lower-level language.
Traditional decompilers, therefore, produce code that is often a shell of its former self, missing many of these abstractions that make source code readable in the first place.
Figure~\ref{fig:intro_example} shows an example of decompiled code relative to the corresponding original source code.

\begin{figure}
\begin{subfigure}[t]{0.50\textwidth}
\begin{lstlisting}[style=cstyle,basicstyle=\ttfamily\bfseries\scriptsize]
void *hash_find(struct hash *h, void *item)
{
    int index = hash_find_index(h, item);
    
    if (index == -1)
        return ((void *)0);
        
    void *out = gap_get(h->data, index);
    
    return out;
}
\end{lstlisting}
\caption{Original Source Code}
\label{fig:intro_original}
\end{subfigure}
\begin{subfigure}[t]{0.49\textwidth}
\begin{lstlisting}[style=cstyle,basicstyle=\ttfamily\bfseries\scriptsize]
__int64 func5(__int64 a1, __int64 a2) {
  __int64 result; // rax
  int v3;         // [rsp+1Ch] [rbp-4h]

  v3 = func4(a1, a2);
  if (v3 == -1)
    result = 0LL;
  else
    result = func1(*(_QWORD *)(a1+8), v3);
  return result;
}
\end{lstlisting}
\caption{Deterministically Decompiled Code}
\label{fig:intro_decompiled}
\end{subfigure}
\caption{A function (\ref{fig:intro_original}), and the same function after being compiled then decompiled (\ref{fig:intro_decompiled}). The decompiled code is missing identifier names, has some incorrect or misleading types, and is structured differently. While it is easy to tell what the function on the left does, it is difficult to tell what the function on the right does. This effect matters because reverse engineers use decompilers to help understand the program under analysis.}
\label{fig:intro_example}
\end{figure}

Neural decompilers~\cite{idioms,llm4decompile,slade,nova,degpt} are machine learning models which help by \emph{predicting} the code in a higher level language.
Neural decompilers may add variable names, reconstruct types, restructure control flow and rewrite code in a more idiomatic way.
Neurally decompiled code can be much more similar to the original source code that was originally used to build the program than deterministically decompiled code.

While neural decompilers are very powerful, they are not perfect: as machine learning models, they offer no correctness guarantees and can make mistakes.
We call an incorrect prediction a \emph{misprediction}.
Particularly consequential are semantic mispredictions: those where the functionality of the program is incorrectly represented.

In this work, we introduce \faultless,\footnote{Code available at https://github.com/squaresLab/faultless} a technique to automatically detect semantic mispredictions.
Using \faultless, a reverse engineer can know whether or not to \emph{trust} the output of a neural decompiler.
We refer to this as \emph{validating} the neural decompiler’s prediction (in the sense of translation validation~\cite{pnueli1998translation,necula2000translation}).
\faultless is a program equivalence technique that compares the output of a neural decompiler with that of a deterministic decompiler; deterministic decompilers produce less readable code but do so with stronger correctness properties.\footnote{Deterministic decompilers are often unsound, but tend to be unsound in fairly predictable ways~\cite{taxonomy,votipka2020observational}. \faultless accommodates these as discussed in Section~\ref{sec:defining_equivalence}.} 
Given two C functions, \faultless will determine whether or not the two are equivalent.

\faultless is also helpful for a related task: evaluation.
Evaluating neural decompilers’ semantic correctness has historically been a difficult part of research into neural decompilers themselves.
Early techniques used heuristics like measuring token overlap or edit distance~\cite{hosseinibeyond,coda}.
\begin{wrapfigure}{r}{0.5\linewidth} % r = right side, 0.5\linewidth = width of image block
    \centering
    \includegraphics[width=0.99\linewidth]{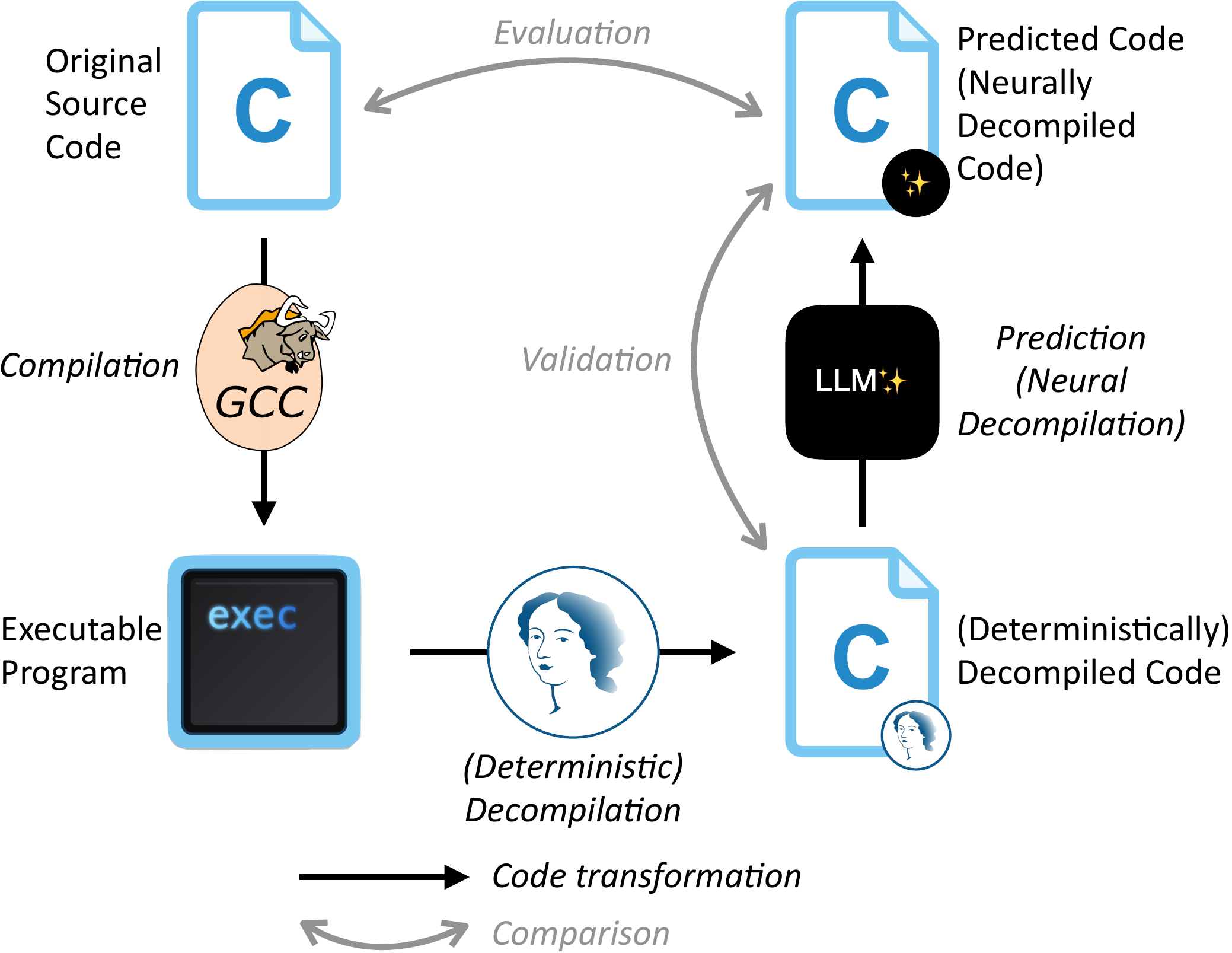}
    \caption{The four code representations in the neural decompilation problem, the relationships between them, and the two comparison tasks that \faultless performs.}
    \label{fig:code_representations}
\end{wrapfigure}
More recently, some techniques~\cite{slade,llm4decompile,nova} have used benchmarks bundled with unit tests: if a neural decompiler’s prediction passes all of the unit tests, it is considered correct.
While better, generating unit tests for arbitrary, sophisticated real-world functions is a very difficult problem, and as a result existing benchmarks tend to be biased towards simpler and easier functions~\cite{idioms}.
Finally, Codealign~\cite{codealign}, released last year, is the first technique specifically designed to evaluate neural decompilers, but its reliance on control and dataflow dependencies makes it very brittle to small, semantically irrelevant differences in the prediction relative to a reference solution.

Validation and evaluation are highly related tasks.
Both are program equivalence problems comparing two code representations that occur in the context of neural decompilation.
In evaluation, the neurally decompiled code is compared with the reference original code.
In validation, the neurally decompiled code is compared with the deterministically decompiled code.
Figure~\ref{fig:code_representations} illustrates these two tasks in relation to the neural decompilation process as a whole.

Program equivalence is a very difficult problem (an undecidable one, in fact) and performing it in the domain of decompilation offers some major additional challenges.
Perhaps the most far-reaching challenge---that which has the greatest impact on \faultless' design, and which also makes most alternative prior work unsuitable for the tasks of validating and evaluating neural decompilers---is the context-independent nature of the problem.
By this, we mean that the functions under analysis are arbitrary functions from inside the program and that most dependency information is not available.
This means, for instance, that when attempting to show these functions equivalent, we don't have access to the calling context, the definitions or declarations of callees, nor the declarations of global variables.
This constraint is realistic in the context of neural decompilation.
Reverse engineers will often do a high-level sub-component scanning step when beginning to work on a binary~\cite{votipka2020observational}; they may choose any part of the binary that looks most promising to being their investigation.
Further, binaries may be quite large and it may not be cost effective, practical or even possible with current techniques to neurally decompile the entire binary, especially in a way that renames symbols consistently across the entire neurally decompiled binary so that these dependencies can be reliably found.

A key implication of this challenge is that the code can't be compiled by typical C compilers which require full definitions of all dependencies.
This aspect of the problem renders many existing program equivalence checking techniques unusable for this problem.
This includes unit tests, because proper compilation is a prerequisite to execution.

One dependency \faultless does rely upon is type definitions.
Some prior work includes type prediction~\cite{idioms}, and LLMs are capable of generating the necessary type definitions (or at least partial definitions) for functions they neurally decompile, as we show in Section~\ref{sec:utility}. 

The validation task also introduces some additional challenges.
Deterministically decompiled code may be semantically nonequivalent to the original, though usually in ways that are well-understood.
For instance, a decompiled function may return a value where the original is \cinline{void} because the compiler uses the return register as a general-purpose scratch register and the decompiler conservatively assumes that the temporary value within should be returned.
Alternatively, a compiler may widen a 32-bit integer variable to a 64-bit one, causing a deterministic decompiler to assume that variable is the larger type.
Neural decompilers may not make the same conservative assumptions as a deterministic decompiler.
Thus, there may be subtle semantic differences between a deterministic decompilation and a neural decompilation, even if the prediction of the neural decompiler is ideal: that is, identical to the original code.
Because of this, a na\"{i}ve application of existing program equivalence techniques to this problem would result in spurious false negatives where the deterministic and neural decompilers' predictions are ``essentially'' the same are flagged as nonequivalent.
Effectively, to be useful, a carefully-callibrated degree of flexibility is required for equivalence checking between deterministically and neurally decompiled code.
To this end, wherever possible, \faultless ignores compiler-specific details and instead tries to capture the \emph{algorithm} that the functions under comparison implement.

It is this property of the problem---that the functions under analysis are by construction implementations of the same algorithm---that offers a key opportunity that help inform one of the main design decisions of \faultless.
Iteration is the property of programs that make program equivalence especially difficult.
At a high level, two implementations of the same algorithm should have the same \emph{loop arrangement}.
That is, for each loop in the original source, there is a corresponding loop in the decompiled code.
Following prior work~\cite{codealign,necula2000translation}, we say that two loops are equivalent iff (1) the state of memory at the start of the loops is the the same (2) if the $i$th loop iterations modify memory in the same way and incur the same side effects.

For the validation task, this may not always hold due to optimizations; some optimizations do restructure loops.
(This is not a problem for the evaluation task because neural decompilers are trained to undo optimizations).
In this context, if using an LLM, it is possible to simply prompt the model to preserve the loop structure of the deterministically decompiled code while otherwise improving the code as much as possible. We explore this possibility in RQ6 (Section~\ref{sec:rq6}).

Our experiments show that \faultless is substantially more capable of performing evaluation and validation of neural decompilers than prior techniques. We also show that \faultless is very efficient, performing 90.1\% of equivalence checks on functions in the realistic \realtype dataset~\cite{idioms} in under one second and 96.7\% in under two seconds.

In short, we contribute:
\begin{itemize}
\item \faultless, an equivalence-checking technique for evaluating and validating neural decompilers.
\item A \href{https://github.com/squaresLab/faultless}{\textcolor{blue}{\underline{python implementation}}} of \faultless totaling more than 11,500 lines of code. \faultless has only two dependencies: \cinline{tree_sitter}, a parsing library, and \cinline{z3}, an SMT solver.
\item A comprehensive evaluation measuring completeness and runtime performance, providing empirical evidence for soundness, and showing how \faultless can be directly applied to neural decompilation with state-of-the-art models.
\end{itemize}

% ------------------------
\section{Related Work}
% ------------------------

Decompilation is a very difficult task, made so by the lossy nature of compilation and the sophistication of the code transformations performed during compilation.
In recent years, there has been increasing interest in applying machine learning to decompilation.
Program equivalence has a longer history, but many existing techniques make assumptions that do not hold in the task of neural decompilation. In this section we outline prior work in neural decompilation and program equivalence.

\subsection{Neural Decompilation}

Neural decompilation approaches attempt to reconstruct the original source code for a particular binary.
Some neural decompilation approaches take assembly as input; others use the output of a deterministic decompiler.
This latter approach is sometimes called decompilation refinement~\cite{llm4decompile}.

Early work in neural decompilation focused on adopting more powerful architectures and tailoring them to the task of neural decompilation.
Katz et al.~\cite{katz2018} use a recurrent neural network.
Coda~\cite{coda} uses a multi-stage model with tree-based encoder and decoder to generate an approximate solution followed by an ensemble of other models to edit the first stage's output.
Cao et al.~\cite{cao2022boosting} use a graph neural network.

Since the transformer~\cite{transformer} architecture has become dominant, work has shifted to scaling these models in size and leveraging them in novel ways.
BtC~\cite{hosseinibeyond} uses a transformer in its encoder-decoder (sequence to sequence) configuration.
SLaDe~\cite{slade} is 50\% larger than BtC~\cite{hosseinibeyond} and uses numerous small adjustments.
LLM4Decompile~\cite{llm4decompile} introduce a family of large causal transformer models with sizes in the billions of parameters.
LLM4Decompile models come in two flavors: one that takes assembly as input and one that takes deterministic decompilation from Ghidra as input.
Nova~\cite{nova} is designed to handle assembly using a custom hierarchical self-attention mechanism designed to handle the low information density and long sequence lengths of assembly code.
\idioms~\cite{idioms} is a neural decompilation technique that produces the definitions for user-defined types used within the code it generates.
This addresses a key gap in prior work, which is underspecified with respect to user-defined types.
To help predict the UDT definitions, \idioms leverages interprocedural context, an approach not seen in prior neural decompilation work.

\subsection{Program Equivalence}

\faultless performs a program equivalence task.
Program equivalence is undecidable; iteration (via loops or recursion) is the concept which makes it so.
Approaches to handling loops can be grouped into four main categories: (1) assuming an iteration in one loop matches an iteration in another~\cite{necula2000translation,symdiff,felsing2014automating,de2016relational,barthe2011relational}, (2) using bounding to limit the number of iterations to a user-specified maximum value~\cite{ardiff,alive2,symdiff,zou2024d}, (3) relying on an external source of additional information, like a human engineer, execution trace, or compiler (during optimization), to help infer loop-related properties~\cite{churchill2019semantic,namjoshi2013witnessing,kanade2009validation}, and (4) using heuristics designed to take advantage of certain properties of the input and output programs~\cite{dahiya2017black,churchill2019semantic,gupta2018effective}.
Early work is primarily based on the syntactic form of the input programs~\cite{yang1989detecting,necula2000translation}, while much more recent work has taken advantage of advances in SMT solvers to decide equivalence~\cite{ardiff,alive2,dahiya2017black,churchill2019semantic}.

One of the most well-studied applications of program equivalence is \emph{translation validation}~\cite{pnueli1998translation}: ensuring that the input and output of a program that changes the syntactic form of a program are semantically equivalent.
The primary use case for translation validation is checking that compiler optimizations do not affect program semantics~\cite{necula2000translation}.
Alive2~\cite{alive2} uses loop unrolling to handle loops, which is unsound (a validated translation may actually be not semantics-preserving), but can still detect many interesting mis-optimizations in LLVM.
It is designed to ensure that no new undefined behavior is added during optimization.
Some compiler validation techniques use information provided by the compiler to help perform optimization~\cite{necula2000translation,kanade2009validation,namjoshi2013witnessing}.
Neural decompilers are black-box machine learning models and thus cannot inform the equivalence-checking process.

Other equivalence techniques are more amenable to black-box code transformations.
Churchill et al.~\cite{churchill2019semantic} use test cases to create traces for the input programs, then compares them to suggest relationships between states in the input programs.
JFTGs~\cite{dahiya2017black,gupta2018effective} bundle control flow with nonbranching code and attempt to match branches in the input programs with each other with a heuristic guess-and-check strategy; Gupta et al.~\cite{gupta2020counterexample} use a similar technique guided by counterexamples.
These approaches either use execution information which is difficult and possibly dangerous to obtain in neural decompilation scenarios or are very computationally expensive.
Another approach is to summarize the entire program with constrained Horn clauses and using an appropriate solver to show equivalence~\cite{felsing2014automating,de2016relational}.
These approaches require the synchronization of one loop with another.
A very different paradigm for translation validation is that of equality saturation~\cite{tate2009equality,stepp2011equality}, which involves repeatedly applying transformations known to be semantics preserving.
This approach is limited by the manually-specified transformation rules.

A related task is \emph{semantic differencing}: finding a \emph{functional} difference between two progams, such as a counterexample generated by an SMT solver.
Differential symbolic execution~\cite{DSE} identifies textual or AST-based differences in code, symbolically executes them, and queries the SMT solver to check their equivalence.
ArrDiff~\cite{ardiff} builds on this work by extracting information from unchanged code blocks that may nonetheless be useful for showing the diffs equivalent.
Symbolic-execution-based approaches are useful, but are unsound with respect to loops, are expensive to run, and must usually compile the input programs, which renders them useless for tasks like evaluating neural decompilers.
In contrast, SymDiff~\cite{symdiff} translates input functions into Boogie~\cite{boogie}, a verification language, and creates logical formulas summarizing the effects of the functions, then uses an SMT solver to check their equivalence.
This approach cannot fully handle loops; they must be unrolled to a specified depth or translated to tail-recursive functions, the latter of which are checked for equivalence separately.

% ------------------------
\section{Faultless}
% ------------------------

\begin{figure}
\includegraphics[width=\textwidth]{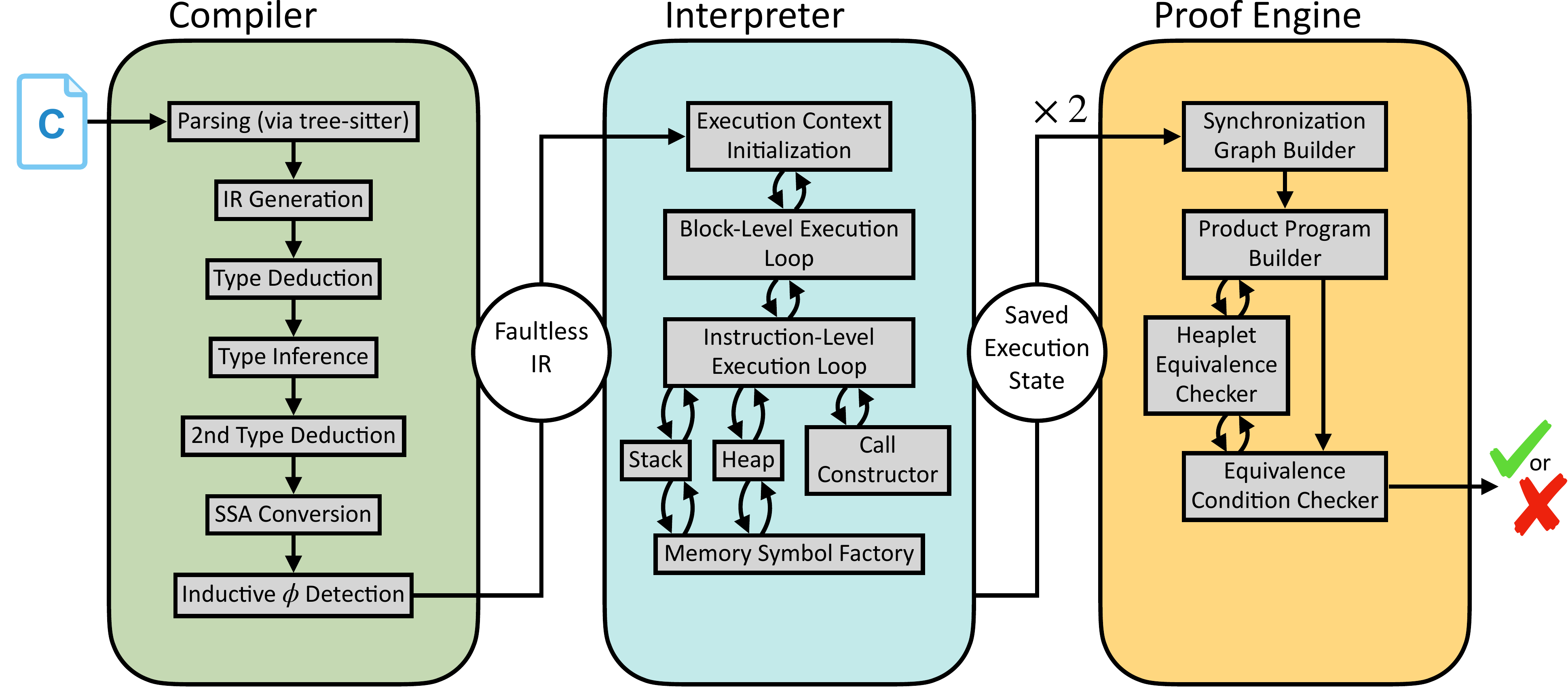}
\caption{An overview of \faultless. It takes as input two C functions and outputs a proof result (equivalent or nonequivalent with a diagnostic reason). It contains three main stages: a compiler, which converts raw C code into a custom intermediate representation (IR), an interpreter, which symbolically executes a function in this IR and records information about the execution, and a proof engine, which uses this information to construct a proof to, if possible, show the programs are equivalent. The first two stages are run twice, once for each of the two functions being compared.}
\label{fig:faultless_overview}
\end{figure}

The \faultless proof system contains three main components: a compiler, a symbolic interpreter, and a proof engine, as shown in Figure~\ref{fig:faultless_overview}.
The technique takes two functions as input---in evaluation, the  predicted (neurally decompiled) code and the reference original code, and in validation, the deterministically decompiled code and the predicted code---and returns a proof result, which can be ``equivalent'' or ``unable to show equivalent.''
(Because \faultless is sound, there may be equivalent functions that it is unable to prove equivalent; these cannot always be distinguished automatically from nonequivalent functions. For convenience, we may refer to an ``unable to show equivalent'' judgement as a ``nonequivalent'' one.)
The first two components are run in sequence for each function.
Then the resulting execution contexts are compared in the proof engine.

In the following section, we'll begin by defining the objective: that is, what we mean by ``equivalence.''
We'll then introduce two core abstractions, the execution model and the memory model, that influence nearly every other component.
Finally, we'll step through each component of faultless in order.

\subsection{Defining Equivalence}
\label{sec:defining_equivalence}

The way \faultless defines equivalence is carefully calibrated to balance utility and a robust notion of equivalent observable behavior.
A C function can return a value, modify program state that persists beyond the deallocation of its stack frame (i.e., the state of the heap or global variables), or read input or output; \faultless captures all of these.
In this section, we discuss the equivalence criteria in an informal way to provide intuition for what \faultless is trying to achieve and to motivate design decisions.
We will formalize these conditions once the execution model, memory model, and other abstractions have been defined.

\faultless uses a broad notion of functional equivalence.
To programs are functionally equivalent if they output the same values given the same inputs.
\faultless' notion of inputs and outputs are not limited to the values of parameters and return values.
\faultless also considers the values reachable from pointers; that is, the values in memory found at the memory address stored in the pointer.
Global variables and function calls can serve as inputs and/or outputs, and \faultless models these as well.

The minimum necessary conditions for \faultless to declare equivalence (hereafter referred to as the ``necessary equivalence conditions'') are, when given the same input:
\begin{itemize}
\item the two functions return the same values.
\item if the return values represent valid memory addresses, then memory on the heap reachable from the return value in each function is equivalent (contains the same values at the same offsets).
\item at the function's exit, memory on the heap reachable from each argument to the function is equivalent to the memory reachable from the corresponding argument in the other function. This captures pass-by-reference behavior. Note that by context independence, \faultless doesn't know  if the pointee referenced a pass-by-reference argument is on the heap or the stack (a fact which may differ by callsite), so without loss of generality \faultless assumes that memory addresses in arguments are locations on the heap.
\item function callees, functions that are called inside the functions whose equivalence \faultless is trying to determine, may also modify program state, read input or write output. Therefore, \faultless requires each callee in one function is equivalent to one in the other. To prove one callee equivalent to another, faultless requires:
\begin{itemize}
\item the arguments to function callees are the same.
\item to capture pass-by-reference behavior to function callees, memory on the heap reachable from each argument is equivalent to memory reachable from the corresponding argument in the other callee.
\item function callees occur under equivalent control-flow conditions.
\end{itemize}
\item function names and global variable names are consistent between the two functions. Neural decompilation may rename global variables and function callee names, so they may not exactly match those in the decompiled code or those in the original code. Instead of requiring them to match exactly, which is often too strict in the context of both the evaluation and validation tasks, we require a bijective mapping to exist between the function names in each input function and likewise for the global variables. We discuss consistency in more detail in Section~\ref{sec:checking_equivalence_conditions}.
\item mapped global variables store the same values.
\end{itemize}

It is also instructive to consider what \faultless does \emph{not} require to show equivalence. 
\faultless intends to capture algorithmic behavior while abstracting away compiler-specific details wherever possible.
Therefore, \faultless may declare two functions equivalent even if they have observably different side channels or may be vulnerable to different specific exploits (e.g. those that depend on a specific stack layout).
There are two other noteworthy areas in which \faultless does not require equivalent behavior or equivalent state: in handling decompiler artifacts during validation, and with regards to side effects.

\begin{table}
\caption{Options \faultless provides to loosen equivalence checking to accommodate common decompiler patterns that result in trivially nonequivalent behavior.}
\label{tab:equivalence_options}
\scriptsize
\begin{tabular}{llll}
\toprule
Code ID & Code Name & \faultless Option Name & Description \\
\midrule
C0.b. & Incorrect type & Math-integer modeling & Model integers as mathematical integers ($\mathbb{Z}$) instead of bit-vectors \\
\hline
C0.b. & Incorrect type & Decompose compound values & In decompiled code, structs in argument lists are often represented \\
& & in parameter lists &  as multiple independent arguments in sequence. Perform the same \\
& & & flattening operation on the original code's structs in argument lists. \\
\hline
C4.a.i. & Extra arguments & Ignore extra arguments & When extra arguments are provided to the function or a callee, \\
& & & assume that they are decompiler artifacts and ignore them. \\
\hline
C12. & Incorrect return behavior & Ignore mixed return behavior & If one function is void and the other is not void, do not compare \\
& & & return values. \\
\hline
C0.c. & Incorrect function name & Require exact function names & Require equivalent functions to have equivalent names. \\
& & & Off by default.\\
\bottomrule
\end{tabular}
\end{table}

As previously noted, decompiled code is often not semantically equivalent to the original source code it was compiled from, though often in very well-understood ways.
To prevent trivial ``nonequivalent'' results, \faultless is equipped with a range of options each of which narrowly loosens the necessary conditions for equivalence, each of which is derived from a code from the codebook in Dramko et al.'s taxonomy of C decompiler fidelity issues~\cite{taxonomy}. These options are summarized in Table~\ref{tab:equivalence_options}.

Additionally, \faultless does not require that side-effect-inducing expressions (such as calls to other functions) be in any particular order. For instance, \faultless judges the following functions equivalent:

\noindent % Prevents paragraph indentation from pushing the layout out of bounds
\begin{minipage}[t]{0.45\linewidth}
\begin{lstlisting}[style=cstyle,basicstyle=\ttfamily\bfseries\scriptsize ]
void foo() {
    printf("a\n");
    printf("b\n");
}
\end{lstlisting}
\end{minipage}%
\hfill % Pushes the second minipage to the right edge
\begin{minipage}[t]{0.45\linewidth}%
\begin{lstlisting}[style=cstyle,basicstyle=\ttfamily\bfseries\scriptsize]
void foo() {
    printf("b\n");
    printf("a\n");
}
\end{lstlisting}
\end{minipage}

The alternative---requiring that side effects do occur in a particular order---would be very limiting.
Due to the context-independent nature of the problem, \faultless does not know which callees and callers have side effects, nor does it know which global variables are read or written from within each callee.
This means that to properly model function calls, \faultless would have to conservatively assume that every function has side effects and that it reads from and writes to every global variable\footnote{Effectively, this means that \faultless assumes that callees don't read or write to global variables.}.
Such assumptions would make the equivalence checking very brittle, limiting the utility of \faultless.

Finally, the current implementation of \faultless has several soundness limitations which mean that functions which produce different outputs for the same inputs may be judged equivalent.
These are discussed further in Section~\ref{sec:limitations:soundness}.

\subsection{Execution Model}
\label{sec:execution_model}

To prove two functions equivalent, \faultless must show that for any input, the two functions return the same value, read the same inputs, produce the same outputs, and modify program state in the same way.
To do this, \faultless uses a symbolic-execution-based approach, providing each function with a set of arbitrary inputs in the form of symbolic variables.
It then incrementally builds up mathematical expressions in terms of these symbolic variables by executing versions of each operation (i.e. \cinline{+}, \cinline{-}, \cinline{sizeof}, etc.) which have been redefined to operate on symbolic expressions instead of the usual concrete values.

Unlike many contemporary techniques, \faultless uses static symbolic execution instead of dynamic symbolic execution.
In symbolic execution, when a branch instruction is encountered, if the value of the conditional is a symbolic expression containing a symbolic variable, execution may fork to \emph{both} the true and false branch.
In dynamic symbolic execution, the two paths are maintained separately in the symbolic execution interpreter, and are both executed to completion.
Usually there is just one copy of the interpreter which alternates between different paths, executing a fragment of one, then another.
Unfortunately there may be a number of paths exponential in the number of branch instructions encountered, a problem known as the path explosion problem.
And because loop branch conditions may be defined in terms of symbolic variables, there may in fact be an unbounded number of possible paths, meaning symbolic execution on any function with such a loop will never terminate.

Static symbolic execution is different in that it features path \emph{merging}.
If two paths would execute the same code, the instead the interpreter halts execution and combines the state of the two paths before proceeding.
This means combining the modifications made to the stack and heap from each path, as well as disjoining the path conditions from each path, signifying that that code fragment could be reached by following one path or the other.

More concretely, \faultless' execution model operates on basic blocks in a control flow graph.
Basic blocks are sequences of instructions with no intervening control flow which serve as nodes in the control flow graph.
A basic block is ready to execute when all of its predecessors in the control flow graph have been executed.
The entry basic block has no predecessors and so is ready to be executed immediately when the interpreter starts up.
Each basic block is executed in sequence with no interruption.
When the basic block is finished executing, its final state---the state of the stack, heap, and path condition to this point---are saved.
Then, for each of that block's successors, the interpreter checks to see if all of the successors' predecessors have been executed and thus have final states saved.
If so, then it merges the incoming states, and adds the merged state and the corresponding basic block to a queue of ``ready'' basic blocks.

\subsubsection{Loops}
\label{sec:loops}

The above algorithm is only guaranteed to progress through the control flow graph if the control flow graph is acyclic.
However, this is not the case for many control flow graphs due to the presence of loops.

To address this problem, \faultless identifies which basic blocks are impacted by identifying loops from the control flow graph using a standard natural loop analysis pass~\cite{reduciblecfgs,dragonbook}.
(Unnatural loops---which in C can only be created with \cinline{goto}s---are not supported.)
\faultless chooses the loop head as the location to break the cycle, because a loop head is always the first basic block in the loop that is reachable from the entry block of the function.
When a loop head is encountered, only the predecessor from the outside of the loop will have a completed end state.
To make up for the lack of other predecessors, \faultless abstracts over all possible values that the variables could take on during the loop, assigning each variable whose value changes in the loop a unique symbolic variable.
\faultless identifies such variables using SSA (Single Static Assignment) form~\cite{alpern1988ssa,rosen1988ssa} $\phi$ instructions, so these variables are referred to as $\phi$-variables.
In SSA-form code, each variable in the program is mapped to a collection of variables, each of which is written to exactly once.
$\phi$ instructions are flow-sensitive copy operations placed at the start of basic blocks that receive a different value for a program variable from different successors.
A $\phi$ instruction at a loop head, then, indicates precisely the program variables that are mutated by the loop and thus must be abstracted over by \faultless with a symbolic variable.
In essence, $\phi$ variables parameterize an arbitrary loop iteration, just as symbolic arguments to the function's parameters represent an arbitrary call to the function.

The introduction of $\phi$ variables mean that symbolic expressions that occur during the execution of the function may be defined in terms of not only the function inputs, but also one or more $\phi$ variables.
This is problematic for some tasks that symbolic execution is traditionally used for, such as finding example inputs that lead to a crash or vulnerable state.
However, it is actually advantageous here, because of the way \faultless performs inductive proofs of loops in the proof engine (Section~\ref{sec:building_the_coupling_relation}).

Because loop bodies can be executed multiple times, a given $\phi$ variable can represent multiple dynamic values.
This stands in contrast to a regular symbolic variable, which represents a singular dynamic value.
A variable that is modified during a loop may be used after that loop terminates, but at that point, the variable also represents a singular dynamic value.
This distinction is important with regard to the memory model.
Therefore, for each $\phi$ variable, we introduce a corresponding post-loop $\phi$ variable, which we denote $\hat{\phi}$.
We refer to ``regular'' $\phi$ variables that model an arbitrary loop iteration as mediloop $\phi$ variables, from the Latin word for ``middle.''

\subsection{Memory Model}

\faultless features a 3-layer memory model consisting of virtual registers, the stack, and the heap.
Registers are the simplest part of the memory model.
They store intermediate values in multi-operation expressions which are not assigned to any particular program variable, like the value \cinline{x * 2} in the expression \cinline{y = x * 2 + 1}.
The \faultless interpreter has an unbounded number of virtual registers.

\begin{figure}
\centering

% Manually create the figure number now, so labels refer to this figure.
\refstepcounter{figure}
\label{fig:address_space}
\setcounter{subfigure}{0}

\begin{minipage}[t]{0.50\textwidth}
\vspace{0pt}

\begin{lstlisting}[style=cstyle,basicstyle=\ttfamily\bfseries\scriptsize]
void func(int *a, int x) {
    a[0] = 5;
    if (x > 0) {
        a[1] = 15;
        a[2] = 25;
        if (x < 3)
            foo(a[x]);
        else
            bar(a[x]);  
    }
    a[x] = 100;
}
\end{lstlisting}

\manualsubcaption{fig:address_space:code}
{A function which writes to `\cinline{a}' at different points and offsets in the control flow graph.}

\vspace{0.5em}

\caption*{\figurename~\thefigure: A function and the address space graph for the heap variable \cinline{a}. If there were more variables written to in \cinline{func}, there would be an address space graph for each one. Address Mapping objects like the Heap and Stack map from the base address to a node in the AddressSpace graph, representing the state of memory at that point in the code.}

\end{minipage}
\hfill
\begin{minipage}[t]{0.47\textwidth}
\vspace{0pt}

\includegraphics[width=\linewidth]{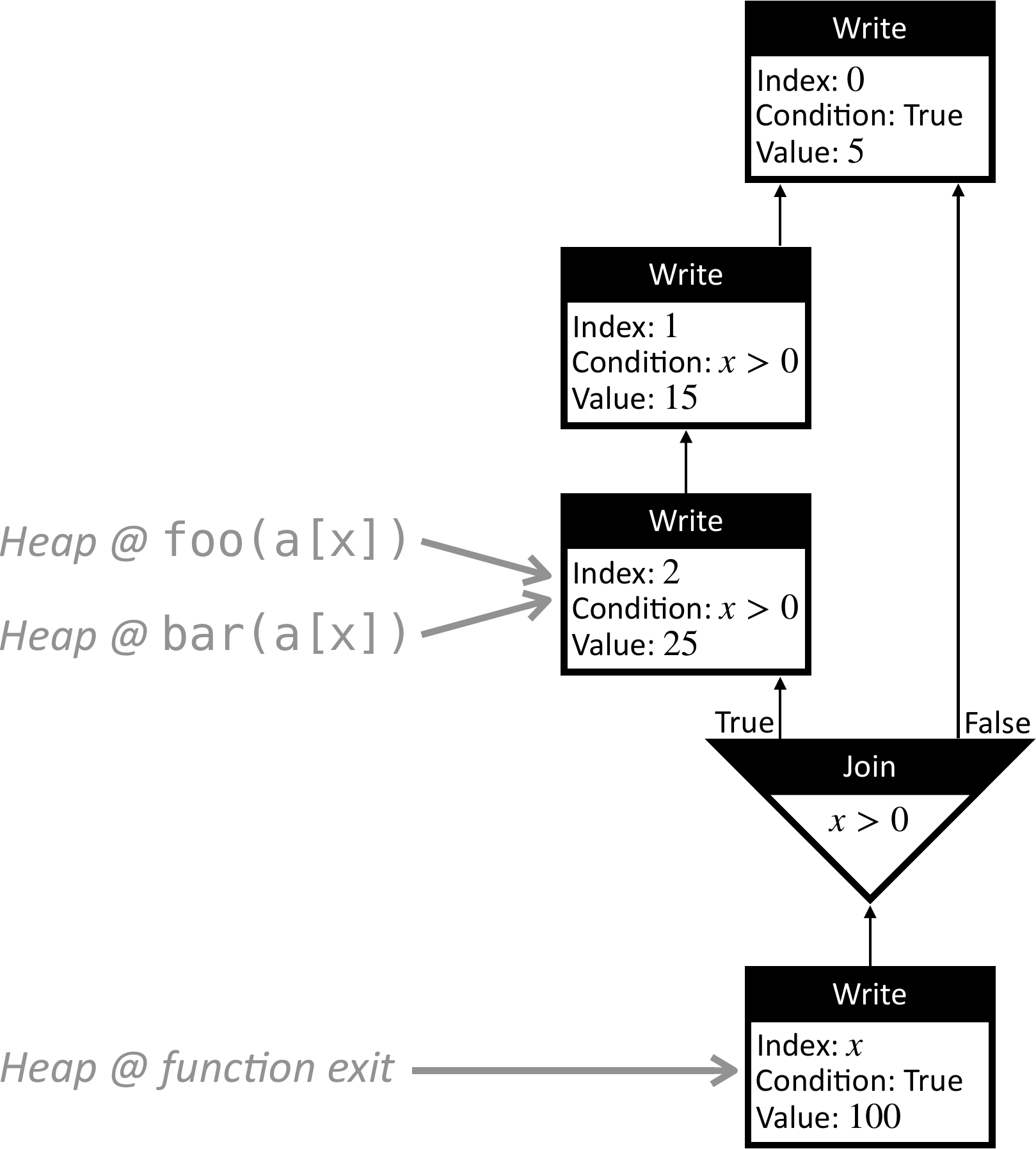}

\manualsubcaption{fig:address_space:graph}
{Heap Address Space Graph for `\cinline{a}'}

\end{minipage}
\end{figure}

Much of the power of \faultless' memory model, however, comes from the stack and heap.
The stack and heap are each instances of an abstract Address Mapping data type.
An Address Mapping is a map from a base address to a segment of memory, similar to a ``heaplet'' in separation logic~\cite{separationlogic}. 
Each memory segment is its own unbounded address space.
The stack is an Address Mapping indexed by program variables (i.e. variables declared in a program that represent memory locations where data is stored), and the heap is an Address Mapping indexed by symbolic-execution-style symbolic variables.
We will write program variables in monospaced font (\cinline{p}), and symbolic variables in italics ($p$).

The values of an Address Mapping's key-value store are themselves an abstract data type, an Address Space.
An Address Space is conceptually a flow-sensitive unbounded symbolic array implementing McCarthy's array axioms.
Each Address Space is implemented as an immutable directed acyclic graph (DAG); Figure~\ref{fig:address_space} shows an example.
Writing to an address space means adding a node to the graph, with a singular outgoing edge pointing to the most recent prior value.
In this way, the graph represents the complete ``history'' of all writes to a given base address.
The Address Mapping stores the pointer to the most recent node in the history graph.
In this way, an Address Mapping can represent a ``snapshot'' of the state of memory at a given point: because the graph is immutable and the pointers in the graph point only to the past, pointing to a given node represents the state of memory immediately after the memory write pointed to was executed.
For instance, in Figure~\ref{fig:address_space}, the state of the Heap saved for checking the necessary equivalence conditions of the calls \cinline{foo} and \cinline{bar} is at the write on line 5, while the state of the Heap at the function's exit is at the write on line 11.

An Address Space is a graph rather than just a linked list because of how control flow interacts with memory.
Different values may be written down different paths.
When a branch instruction is encountered each resulting path receives a copy of the stack and heap.
The mappings can be mutated independently, adding new nodes which point to the most recent prior node and creating a ``fork'' in the memory graph.

Path merges create an additional challenge because there are multiple possible write histories.
To handle this, Address Space graphs include another type of node, a Join node.
Join nodes are parameterized by a boolean condition and point to two prior history nodes, one for the true branch and one for the false branch.
The boolean condition used is the boolean condition found in the branch which originally caused the divergence in control flow.

\subsubsection{Reading}
While a write simply involves appending the value written to the Address Space graph, it is not immediately clear how to \emph{read} a value from this graph.
To do this, we first need to examine what is stored in each Write (non-Join) node of the graph.
Each Write node contains three attributes:
\begin{itemize}
\item The \emph{index}, relative to the base address, at which the value is stored. For a scalar variable this will simply be 0, but it could be any symbolic expression for an array variable (e.g. for the expression \cinline{a[x + 1]}, the index would be $x + 1$, or, more precisely, the symbolic value stored on the stack in the program variable \cinline{x}, which we can't know without knowing the preceding code, plus one).
\item The path condition at which this write occurred. We call this the \emph{write condition}.
\item The value written.
\end{itemize}
\noindent
An address space represents a conceptually unbounded chunk of memory, with each storage location represented by a unique ID, or address, similar to virtual memory implemented in modern operating systems.
Unlike in virtual memory, the unique IDs are represented as offsets relative to the base address, starting from zero.

A read query is parameterized by an index, which may be symbolic, and the path condition at which the read occurs, the \emph{read condition}.
To read a value, the memory graph is searched, starting from the node pointed to by the Address Mapping for that base address.
If the indices are equivalent under the read and write conditions, then the value stored in that write node should be returned.

That is the ideal, anyway, but unfortunately things do not always work out this cleanly.
The property of reads that makes reading difficult is that indices may be arbitrary symbolic expressions.
While a concrete value like $0$ or $8$ represents a fixed offset with respect to the base address, a symbolic offset like $i$ may be any offset, so, depending on the value that $i$ takes on, this value may or may not be read.
For instance, if we previously executed \cinline{arr[0] = 7} before executing the read \cinline{arr[i]}, (under empty read and write conditions) the index $i$ could take on the value 0, or it could not, so the read value could be 7, but may not be.
However, under the read condition $i > 0$, it is not possible for $i$ to be $0$, and it is therefore impossible for the read value to be 7 (at least from this Write node).
And of course if the read index was also $0$ instead of $i$ (reading back \cinline{arr[0]}) then the read value would be guaranteed to be 7.

\begin{figure}
\begin{minipage}{0.34\textwidth}
\includegraphics[width=\textwidth]{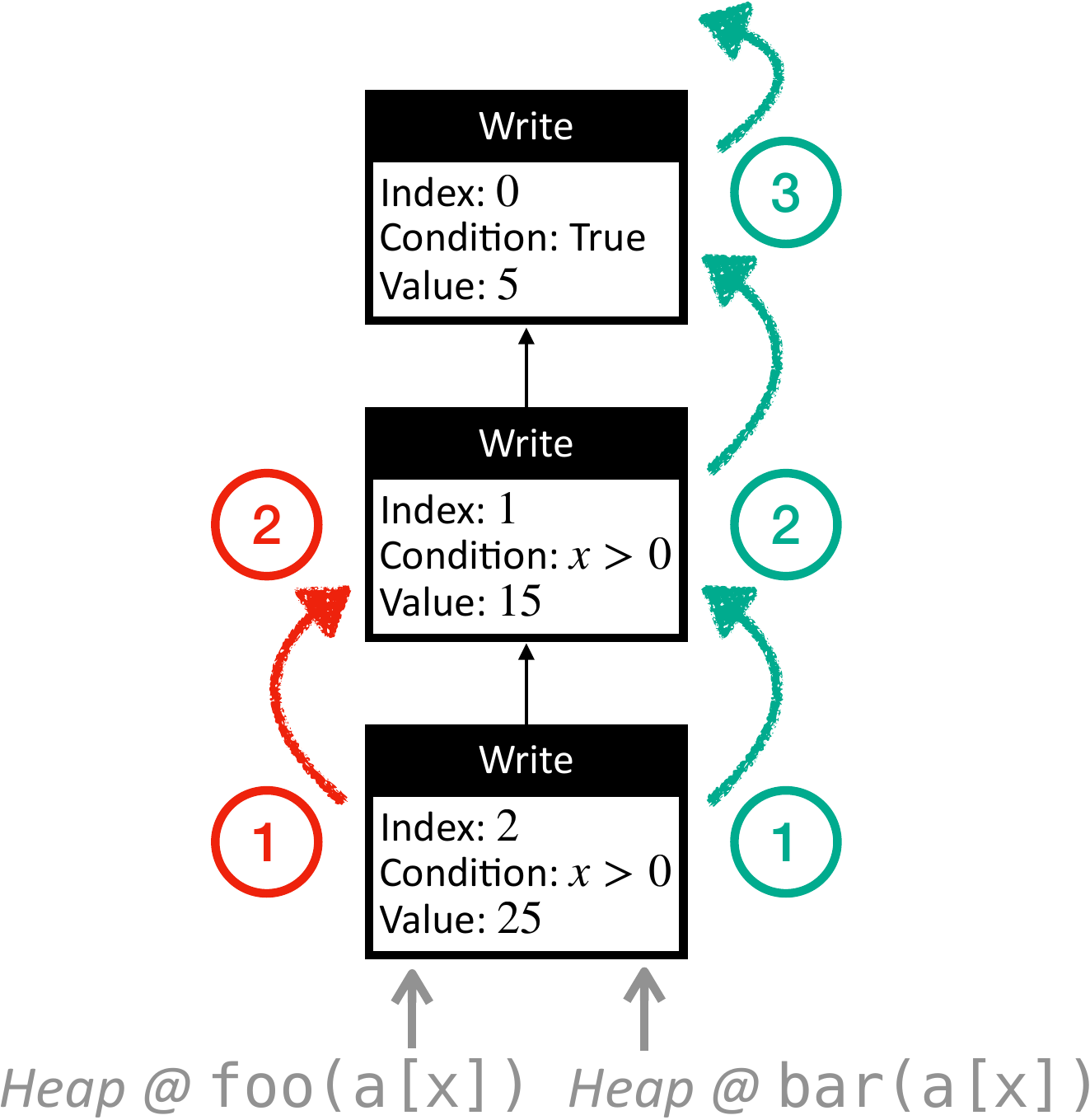}
\end{minipage}
\begin{minipage}{0.65\textwidth}
\centering
\begin{center}
\cinline{a[x]} at \cinline{foo(...)}
\end{center}
{\scriptsize
\begin{tabular}{llll}
\toprule
& Formula & Expression & Result \\
\midrule
\multirow{2}{*}{\color{diagramred}{1}} &
    overlap & $x = 2 \wedge x < 3 \wedge x > 0$ & satisfiable \\
 & coverage & $x < 3 \wedge x > 0 \implies x > 0 \wedge x = 1$ & invalid \\
\midrule
 \multirow{2}{*}{\color{diagramred}{2}} &
   overlap & $x = 1 \wedge (x \le 0 \vee x \ne 2)\wedge x < 3 \wedge x > 0 \wedge x > 0$ & satisfiable \\
 & coverage & $(x \le 0 \vee x \ne 2) \wedge x < 3 \wedge x > 0 \implies x > 0 \wedge x = 1$ & valid \\
\bottomrule
\end{tabular}

Read value: $if(x = 2, 25, 15)$
}
\begin{center}
\cinline{a[x]} at \cinline{bar(...)}
\end{center}
{\scriptsize
\begin{tabular}{llll}
\toprule
& Formula & Expression & Result \\
\midrule
\color{diagramteal}{1} & overlap & $x = 2 \wedge x > 3 \wedge x > 0$ & unsatisfiable \\
\hline
\color{diagramteal}{2} & overlap & $x = 1 \wedge x > 3 \wedge x > 0$ & unsatisfiable \\
\hline
\color{diagramteal}{3} & overlap & $x = 0 \wedge x > 3 \wedge x > 0$ & unsatisfiable \\
\bottomrule
\end{tabular}

Read Value: $d_{\text{\cinline{a[x]}}}$ (a derived symbolic variable (Section~\ref{sec:derived_variables}))
}
\end{minipage}
\caption{A representation of the reads that occur in the argument lists of \cinline{foo} and \cinline{bar} in Figure~\ref{fig:address_space:code}. Both read \cinline{a[x]}, from the same heap, but the differing read conditions cause the two reads to return drastically different values. To perform a read, \faultless traverses the Address Space graph, evaluating coverage and overlap formulas for satisfiability and validity, respectively. A valid formula is one that is always true. In practice, the symbolic value $p_1$, representing the value to the first parameter, would stored in \cinline{x}, but we use $x$ here for clarity.}
\label{fig:example_read}
\end{figure}

To handle this complexity, \faultless introduces two concepts: overlap and coverage.
If a given read query \emph{overlaps} the write, that means there is at least one index where the read and write condition are equivalent, subject to the read and write conditions.
If a given write query \emph{covers} the read, that means that all indices in the read query have a matching index in the write.
There are three possibilities:
\begin{itemize}
\item no overlap: this write cannot be the read value. Search deeper through the write history. It is not necessary to check coverage.
\item overlap but no coverage: this write may contribute to the read value, but we also need to find other possibilities in the write history.
\item coverage: the read value is the value stored at this Write node.
\end{itemize}
Figure~\ref{fig:example_read} contains examples of all three situations; steps {\color{diagramteal}1}, {\color{diagramteal}2}, and {\color{diagramteal}3} for \cinline{bar(a[x])} are no overlap read-write interactions, step {\color{diagramred}1} for \cinline{foo(a[x])} represents an overlap but no coverage interaction, and {\color{diagramred}2} for \cinline{foo(a[x])} represents a coverage interaction.

When there is overlap but no coverage, multiple different read values are possible.
\faultless uses symbolic ``if-then-else'' expressions (\cinline{z3.If}) to represent them as a single value.
The condition of the if-expression is a symbolic formula which captures the indices read from in this read/write iteraction, the consequence is the value of the current Write node, and the alternative is the result from recursively calling read on the next node in the Address Space write history graph.
The recursive call includes an additional constraint, the negation of the if-expression's condition, that prevents recursive read calls from matching over the same indices that overlap the current (and thus more recent) write, progressively refining the query to a smaller set of indices.

Let $\kappa_r$ be the read query index, and let $\kappa_w$ be the write query index. Let $\pi$ be the read condition, and let $\omega$ be the write condition.
All are symbolic formulas.
We discharge queries over symbolic variables to the \cinline{z3} SMT solver~\cite{z3}.
The formulas for overlap, coverage, and the recursive constraint are as follows:

\vspace{4pt}
\noindent 
\emph{Overlap.}
For a write to overlap the read, subject to the read and write conditions, there must be at least one satisfying assignment of values to variables that makes the read and write indices equivalent and for which the read and write conditions are both true.
Thus, the overlap formula is a simple conjunction of the equality of the indices and the read and write conditions.
Formally, there is overlap if the following formula is satisfiable:
\begin{equation}
\label{eq:overlap}
(\kappa_r = \kappa_w) \wedge \pi \wedge \omega
\end{equation}

\vspace{4pt}
\noindent
\emph{Coverage.}
For a write to cover a read, every possible index specified in the read query must also exist in the write.
Read indices are subject to the read condition, so we need only consider the cases when the read condition is true.
But otherwise, for every possible value of the variables in the read and write index formulas, the read indices must be equivalent, and the write condition must allow for them to be.
Formally, the write covers the read if and only if the following formula is valid, that is, always true for any assignment of values to symbolic variables:
\begin{equation}
\pi \implies (\omega \wedge (\kappa_r = \kappa_w))
\end{equation}
This simple form of the equation works in loop-free code.
However, mediloop $\phi$ variables represent multiple dynamic values at once, as opposed to regular symbolic variables, which only represent one.
This means we must quantify over mediloop $\phi$ variables.
Let $\phi_{r_0} \dots \phi_{r_n}$ be the mediloop $\phi$ variables that occur in the read index or read condition symbolic formulas, and let $\phi_{w_0} \dots \phi_{w_m}$ be the mediloop $\phi$ variables that occur in the write index or write condition symbolic formulas.
(We do not have to be concerned about this for overlap because SMT expressions are implicitly existentially quantified and overlap is by its nature purely existential).
The formula says: for all possible mediloop read $\phi$ values that satisfy the read condition, there exists a combination of corresponding write mediloop $\phi$ values subject to the write condition such that the read and write indices are equivalent.
Formally, this is:
\begin{equation}
\label{eq:coverage}
\forall \phi_{r_0} \dots \phi_{r_n}.\; \pi \implies (\exists \phi_{w_0} \dots \phi_{w_m}.\; \omega \wedge (\kappa_r = \kappa_w))
\end{equation}
Because validity is the dual of satisfiability and SMT solvers find satisfying assignments of values to variables, the ask the solver if the negation of Equation~\ref{eq:coverage} is unsatisfiable.

\vspace{4pt}
\noindent
\emph{If-Condition.}
The consequence (the ``true'' branch of the symbolic ``if'' expression) of an overlapping but uncovered read is the value stored in the write node.
The condition of the if-expression, then, must be true for all values for which the read and write indices are equivalent, and false when they are not.
This is similar to what the coverage condition, which Equation~\ref{eq:coverage} expresses.
However, we do not include the read condition.
By construction, the read condition holds over the entire value read.
Including the read condition in the if-expression condition would provide an interpretation of the read value where the read condition is not met, which is nonsensical.
Formally, the if-condition is:
\begin{equation}
\label{eq:ifcondition}
\exists \phi_{w_0} \dots \phi_{w_m}.\; \omega \wedge (\kappa_r = \kappa_w)
\end{equation}
The negation of this formula helps refine recursive queries.
Intuitively, new writes to the same address in memory should overwrite older ones at that same address.
Equation~\ref{eq:ifcondition} formula describes all addresses that are read from the current Write node.
We want to ensure that subsequent reads do \emph{not} read from those indices, so we assert the \emph{negation} Equation~\ref{eq:ifcondition} in conjunction with the read condition.
In step {\color{diagramred}2} for \cinline{foo(a[x])} in Figure~\ref{fig:example_read}, the refinement is the negation of $x > 0 \wedge x=2$, which is $x \le 0 \vee x \ne 2$.

\emph{Join-Nodes.}
Join nodes represent places in the program where write history to memory differs depending on the path taken.
To read from a join node, \faultless also uses an if-expression.
\faultless recursively reads down both the true and false branch, asserting the branch condition and its negation, respectively, as additional refinements.
The if-expression is parameterized by the branch condition, with the result of the recursive call on the true successor of the Join node assigned as the if-expression's true argument and the result of the recursive call on the false successor of the Join node assigned as the if-expression's false argument.
If the Join-node's branch condition is unsatisfiable under the read condition then instead only the value from the false branch's recursive call is returned. Likewise, if the negation of the Join-node's branch condition is unsatisfiable under the read condition, then only the value from the true branch's recursive call is returned.

\subsubsection{Context Independence, Progenitor Variables, and Derived Variables}
\label{sec:derived_variables}

Suppose that there is no write in the write memory graph which completely covers the read for a given query.
In the simplest case, this occurs when a read happens before any write to a given address, such as in the snippet
\begin{lstlisting}[style=cstyle,basicstyle=\ttfamily\bfseries\scriptsize]
void foo(int* ptr) {
    int first = ptr[0];
    // ...
}
\end{lstlisting}

In such a case, it is not clear what value the read should return.
\faultless' answer to this problem lies in its context-independent nature.
By being context-independent, \faultless assumes nothing about the calling context (nor the definitions of callees or global variables).
Thus, \faultless knows nothing about the values in the chunk of memory that the argument \cinline{ptr} is pointing to, nor does it even know the size of that chunk of memory.
Instead, \faultless \emph{records the assumptions} that the function makes about the calling context---in this particular instance, about the calling context memory.
It allocates a fresh, unique symbolic variable for each combination of base address and accessing index, symbolizing that the function assumes that there is a value at that location (or those locations if the index is a symbolic expression capable of representing multiple values).

We call symbolic variables create this way \emph{derived variables}.
Derived variables are derived from other derived variables (as in \cinline{ptr[0][0]}) or \emph{progenitor variables}.
Progenitor variables serve as the roots of derivation trees.
They come from five different sources:
\begin{itemize}
\item symbolic parameters. As mentioned in Section~\ref{sec:execution_model}, \faultless provides a set of symbolic parameters to each function to begin execution. The variable derived from \cinline{ptr} above is derived from a symbolic parameter.
\item $\phi$ variables, both mediloop and postloop, as described in Section~\ref{sec:loops}.
\item function calls. Calls to other functions are treated as uninterpreted, in keeping with the context-independent assumptions of \faultless, and each function call is associated with a unique symbolic value. For more information about function calls, see Section~\ref{sec:function_calls}.
\item global variables. By context independence, \faultless know nothing about the values in global variables when the function is called so each global is initialized with a unique symbolic variable.
\item uninitialized local variables. \faultless initializes all parameters with symbolic values, and most other local variables are initialized with symbolic expressions defined in terms of those variables or with constants. But variables which are read before they are initialized get their own unique symbolic values.
Unlike with the other three types of derived variables, we know what is present in uninitialized locals, or, rather, that they contain no particular value.
\end{itemize}
The majority of the work that \faultless' proof engine (Section~\ref{sec:proof_engine}) does involves mapping progenitor variables to each other so that symbolic variables in expressions across different functions are defined in terms of the same variables.
Once progenitor variables are mapped together, their derived values can be mapped together as well based on the corresponding index expressions.

Derived symbols are created and stored by the Memory Symbol Factory in the interpreter, as shown in Figure~\ref{fig:faultless_overview}.

\subsubsection{The Stack and Compiler Independence}

The stack is simply an Address Mapping indexed by program variable.
Each program variable is mapped to its own Address Space.
As noted above, each Address Space represents an unbounded set of addresses, with the base address mapping to index 0.
This choice may seem odd for modeling a frame on the call stack, since we know the types of all local variables and thus the amount of space each local variable takes up.
This stands in contrast to the heap, for which, by context independence, we know nothing of the size of the memory allocated at each base address so we cannot bound heap Address Spaces by any particular concrete value.

However, using an Address Space model is helpful for the Stack for different reasons.
In addition to being context-independent, \faultless is, wherever possible, compiler-independent as well.
That means \faultless makes no assumptions about the layout of the stack frame, such as the order of variables in the frame or the amount of padding placed between them.
The nature of Address Mappings and Address Spaces allows for this compiler-independent behavior.

For the vast majority of regular local variables, though, all reads and writes just occur at index 0, so each write covers each read and the value read is simply the last value written.

\subsubsection{Addressability}

Both reads and writes to memory happen relative to some base address.
Unlike in normal program execution, these base addresses are not merely integers but are instead symbolic values (for the heap) or program variables (for the stack).
\faultless determines what the base address is for a given expression through its \emph{addressability model}.

A key challenge in modeling addressability comes from the validation task, in which \faultless must compile and execute decompiled code, which features frequent integer-to-pointer casts.
For instance, a frequent occurrence in code produced by the Hex-Rays decompiler is a parameter typed as a \cinline{long long} or \cinline{__int64} which is then cast to a pointer type and dereferenced.
The variable \cinline{a1} in Figure~\ref{fig:intro_decompiled} is an example of this.
A memory address, is, after all, just an integer identifying a particular virtual memory cell; pointers wrap a layer of abstraction around memory addresses, modeling how much memory is being referred to starting from that cell and how to interpret the bit pattern found in those memory cells.
Decompilers simply sometimes strip variables of that semantic wrapping and then manually apply it later.
Thus, we must allow integers to serve as legitimate base addresses as well.
In fact, \emph{any} integer or pointer symbolic value could be a valid base address: by context independence, \faultless doesn't know whether a given symbolic variable is a memory address or not based on the calling context, so if the function uses the symbolic value as if it is a memory address, \faultless records the assumption by treating it as one.

The addressability model is most directly derived from the properties of memory addresses in position-independent code.
\faultless determines the base address not only from the type but from \emph{how each value is used}.
Because memory addresses in position-independent code may vary from execution to execution, only certain operations on valid memory addresses preserve the property that the result may be a valid memory address on an arbitrary run.
For instance, multiplying an address by two will result in a value that is likely outside of the address space to or from which the process can read or write.
Instead, in position-independent code, valid memory addresses can only be obtained by adding or subtracting offsets relative to an existing address.

Addressability is thus a recursive property.
Plain symbolic valariables form the base case.
Then if a symbolic expression is addressable and it is added or subtracted from another symbolic expression, the resulting expression is also addressable.
There is a caveat here: if \emph{both} of the operands to an addition are addressable expressions, what is the resulting base address?
After all, it could be that both operands represent memory addresses, and summing or differencing two memory addresses also leads to an invalid address.
In this case, \faultless takes a hint from the operands' types: if one is a pointer and the other an integer, then the pointer is the base address.
If both are integers (both pointers is a compilation error) then the resulting expression is not addressable because the base address cannot be determined.
Attempting to store a value at a non-addressable lvalue is an execution error.

Supporting integer-to-pointer casts as a first-class feature is both extremely important for supporting decompiled code but also relatively unusual in symbolic execution and equivalence checking techniques; for instance, Alive2~\cite{alive2} explicitly disallows integer-to-pointer casts, and Klee~\cite{klee} has very limited support for them.

\subsection{The Compiler}

\faultless' compiler converts each C function into a high-level three-address-form intermediate representation (IR).
To preserve compiler independence, the level of abstraction is kept as high as possible, preserving source-level details.
In \faultless, compilation, then, serves less as a method of conforming the software to the hardware and more of a way of providing a distinct partial ordering of operations for execution, breaking down large expressions into smaller ones with intermediate results being stored in temporary variables.
The control flow graph is built directly as a first-class feature of IR generation instead of inferred later from branch instructions and their targets in the code.

The \faultless compiler module is based upon the compiler model from codealign~\cite{codealign}, though it features numerous improvements, including the ability to parse and model types. (Codealign is type agnostic.)
We modified the type-parsing code from the \idioms project~\cite{idioms} and incorporated it into the compiler to help accomplish this.

\vspace{4pt}
\noindent
\emph{Context Independence}.
Because \faultless does not require the definitions of callees or globals, it must determine which identifiers correspond to these constructs, which it does through a collection of heuristics, inherited from codealign~\cite{codealign}.
If an undeclared identifier is used as the name of a function in a call expression, we assume it is a function.
If an undeclared identifier is used as a value, it's assumed to be a global variable.
If the definition of a global variable or a function is given, \faultless will parse and use it, making it unnecessary to infer that particular identifier's identity.
\faultless does require the definitions of \emph{types} used in the function, though some existing techniques can provide them~\cite{idioms}, and as we show in Section~\ref{sec:utility}, modern LLMs can be leveraged to generate them as well.

\subsubsection{Type Deduction}
\label{sec:type_deduction}

\faultless applies C's rules around integer promotions, arithmetic type conversions, implicit casts, and assignment semantics to derive the types of intermediate temporary variables and to typecheck the code.
We refer to this process as \emph{type deduction} to distinguish it from the process of \emph{type inference}, described in the next section.

Deterministic decompilers may not be strictly compliant with the C standard's typing rules, which can lead to trivial typing errors when compiling decompiled code for the validation task.
In particular, they often feature size-based placeholder types, like Hex-Rays' \cinline{_QWORD} and Ghidra's \cinline{undefined8}.
The \faultless API features a way to specify which identifiers correspond to these types and their sizes.
They're treated as supertypes of all primitive types of the same size.

\subsubsection{Type Inference}

When \faultless infers that a given identifier is a global variable or function, it does not inherently know that variable's type, and it is initially given a special \cinline{UnknownType}.
But types are necessary to define the functions' semantics.
As a result, \faultless features a module that infers the types of these symbols based on the context in which they are found in the function.

The C programming language's typing rules only provide unique types when applied in the forward direction.
For instance, if \emph{g} is a global variable, we can conclude from the snippet \cinline{short x; g = x + 1;} that \cinline{g} has type \cinline{int} by C's integer promotion rules.\footnote{Actually, \cinline{g} could still be any integer type or even a float type because C will implicitly cast this integer-typed x + 1 expression to the declared type of the variable. But from a type inference perspective, the typing rules do provide an obvious unique type in this case.}
In contrast, there are multiple possible types when applying the rules backwards: in \cinline{int x = g + 1;}, \cinline{g} could be a \cinline{char}, \cinline{unsigned char}, \cinline{short}, \cinline{unsigned short}, \cinline{int}, or even \cinline{_Bool}.

The challenges associated with inferring types for C code are well studied, including for use in decompilers themselves~\cite{retypd,binsub,tie,tiesurvey} and for C code more generally~\cite{psychc1,psychc2}.
\faultless' type inference module was designed in collaboration with and synthesized by AI based upon common patterns in related work.

The type inference module assigns a type variable to each unknown type, which is associated with a domain, a set of possible types that the variable could be.
The inference module generates constraints over those type variables based on their usage.
For instance, if a variable with an unknown type is used in a bitwise \cinline{&} operation, it must be an integer type.
If a variable with an unknown type is dereferenced, it must be a pointer type (though more evidence is required to determine the pointee type).
Applying a constraint to one or more type variables narrows the type domain.
The constraint solver uses a worklist algorithm to propagate constraints, using information in a given constraint to narrow the domains of type variables used in the constraint.
Whenever a variable's domain reduces in size, constraints mentioning that variable are added to the worklist.
The process proceeds to a fixed point.

If a type variable's domain is empty, then type inference has detected an inconsistency in the typing of the input function.
However, it is entirely possible and indeed common for a type variable's domain to contain more than one type at the end of type inference; these represent the set of types that are consistent with the typing evidence available.
This is the case for the \cinline{int x = g + 1;} example from earlier (assuming that \cinline{g} is not used elsewhere in the function).
To handle this, the type inference module maintains a set of preferences associated with each type choice.
Among integer types, for instance, the smallest type that can hold the value without loss of information is preferred.
If there is not enough information to determine this, then \cinline{int} is chosen.

After type inference completes, type deduction is run again to propagate the inferred types of global variables and function signatures through all of the program's temporary variables.

\subsubsection{Preparing Loops}

To further prepare the \faultless IR, it is converted into SSA form.
The interpreter directly executes the SSA form IR; the $\phi$ instructions at loop heads are important for \faultless' abstracted iteration summary execution (described in Section~\ref{sec:execution_model}).
We refer to these $\phi$ instructions as \emph{loop-$\phi$ instructions}.
To differentiate loop-$\phi$ instructions form other $\phi$ instructions, \faultless performs a standard loop-analysis pass, then merges together all natural loops with the same head.
(Natural loops are uniquely identified by back edges; two natural loops can have the same head if a \cinline{continue} statement is used, for instance.)
The operand of each loop-$\phi$ which does not come from within the merged loop body is recorded as the base-case operand.

\subsection{The Interpreter}

The interpreter implements the execution model (Section~\ref{sec:execution_model}) for \faultless IR.
It builds expressions in terms of symbolic variables, and logs information around the return state and function calls for use in the proof engine.

The interpreter initializes the stack by writing position-dependent symbolic variables to each local variable on the stack.
Here we'll denote them $p_0, \dots, p_n$.
Each global variable is also initialized with a unique symbolic variable based on its name.

\subsubsection{Interactions with Memory}

The interpreter directly executes the SSA-form IR.
Each instruction which produces a result is assigned a unique variable; each variable is assigned a register.
Registers are implemented as a simple dictionary mapping an instruction to the result of executing that instruction.

If an instruction's result corresponds to a declared program variable instead of a compiler-generated temporary variable, then the result of executing that instruction is also written to the stack.
Likewise, when an instruction's SSA variable is used as an operand to another instruction, if that instruction wrote to a variable on the stack, that value is read from the stack instead of from the registers.
This is for two reasons.
First, unlike temporary variables, declared program variables can have their addresses taken and can be read from and written to by reference.
Second, the stack already features logic for managing path merges, so ensuring that values are stored there obviates the need for a separate path-merging implementation for registers.
An implication of this second point is that $\phi$ instructions which are not loop-$\phi$ instructions are ignored.

Struct values are written to memory by writing each field to the stack or heap, at the offset given by the ABI for that struct's layout.
This partially breaks compiler independence, or at least platform independence (currently it is implemented for x86\_64) and is the reason \faultless requires UDT definitions.
However, this approach is important for processing decompiled code, where struct-field access operations are often decomposed into arithmetic with a hardcoded offset, a typecast to a pointer, and a dereference, as in Figure~\ref{fig:intro_decompiled} (line 9).
Mapping back to numeric offsets helps ensure that values written to the same fields end up in the same offsets in memory across both the decompiled and original versions of the code in the validation task.
An alternative approach would be to treat field names as symbolic variables representing an unknown offset, but these would have to be carefully constrained to take on consistent values across various memory interactions and to prevent two different fields from being assigned the same offset.

\subsubsection{Executing Function Calls}
\label{sec:function_calls}

Consistent with context independence, \faultless treats function calls as uninterpreted functions.
Each semantically distinct set of function arguments is assigned a unique symbolic variable.
Because function calls may return pointers to memory, these symbolic variables serve as progenitor variables (Section~\ref{sec:derived_variables}).

\vspace{4pt}
\noindent
\emph{Modification-By-Reference Symbolic Variables}.
C functions may also have pass-by-reference behavior, meaning they may return results by modifying memory accessible from the input arguments.
To model this, \faultless features the option to write to the base address of each addressable argument a unique modification-by-reference symbolic variable.
This is a fairly disruptive and invasive option, so it is disabled by default; most functions do not modify any parameters by reference and those that do often only change one or two.
Due to context independence, \faultless cannot know which function arguments are modified by reference.

\vspace{4pt}
\noindent
\emph{Stack Initializer Inference}.
There is one instance where modification-by-reference is especially common: when initializing local variables through functions like \cinline{fscanf} and \cinline{memcpy}.
\faultless also features an option to enable the stack-initializer-inference heuristic: if a local variable is uninitialized and its address is passed to a function call as an argument, that function argument is inferred to be an initializer argument.
Under this option, only initializer arguments are assigned their own unique modification-by-reference symbolic variables.
This option is on by default.

\subsubsection{Entering and Exiting Loops}

Under the execution model (Section~\ref{sec:execution_model}), loops are executed once, with values changed during the loop abstracted to mediloop $\phi$ variables to represent an arbitrary loop iteration.
Mediloop $\phi$ symbolic variables have special semantics in that they represent multiple possible dynamic values instead of a single, arbitrary dynamic value like a regular symbolic variable, so \faultless must convert these to distinct postloop $\hat{\phi}$ symbolic variables after the loop: if the corresonding program variable is read from after the loop, we would expect the value stored there to be in terms of its postloop symbolic variable.

To accomplish this, \faultless ascribes special semantics to loop-$\phi$ instructions.
Each loop head is executed twice, once before and once after the execution of the loop's body; different semantics are given to the instruction on the first and second executions.
Conveniently, all program variables whose values change during the loop have a $\phi$ instruction at the loop's head, and $\phi$ instructions are placed at the start of the basic block, before all other instructions.
Therefore, on the first execution, $\phi$ instructions simply write the corresponding mediloop $\phi$ variable to the stack location for the corresponding program variable.
On the second execution, the $\phi$ instruction writes the postloop $\hat{\phi}$ symbolic variable to the same stack location.
(Loop $\phi$ instructions have additional semantics, which will be covered in the next two sections).

\faultless' compiler emits two types of branch instructions: \cinline{Loop} and \cinline{If} branch instructions.
\cinline{Loop} branch instructions are critical to facilitating this process.
The first time a \cinline{Loop} branch is executed, only the true-branch successor is marked as ready-to-execute.
Thus, the stack initialized with mediloop $\phi$ variables is used for the loop body.
On its second execution, a \cinline{Loop} branch instruction's false branch will be marked as ready-to-execute.
By this point, the $\phi$ instructions will have written postloop $\hat{\phi}$ instructions to the stack, and so these are the exposed values that can be read by code after the loop for those program variables.

Loops may not be exited by just the regular looping condition, however; \cinline{break} statements can also be used to exit the loop, as can \cinline{return} statements, implicitly.
At branches that exit the loop created by these statements, \faultless implements a change-of-variables, reading the current value of each loop-impacted program variable, substituting each instance of the mediloop $\phi$ variable for the postloop one, and writing the resulting value back out to memory.
This change-of-variables is also performed on the path condition.

\subsubsection{Pre-Execution and Loop Invariants}

To better characterize loop iterations, where possible, \faultless infers loop invariants for what we refer to as \emph{generalized affine loops}.
Let $c$ be the change to the value stored in a given program variable \cinline{v} during the execution of an arbitrary loop iteration.
When a value does not change during the loop, $c$ is trivially nothing, so we concern ourselves only with program variables that are modified during the loop, which are precisely those that we initialize with mediloop $\phi$ variables.
Then $c = \text{stack.read}(\text{\cinline{v}}) - \phi_{\text{\cinline{v}}}$, where the stack in question represents the state of memory after the arbitrary loop iteration.
\faultless' inference code supports generalized affine loops: those for which $c$ is not a function of $\phi_{\text{\cinline{v}}}$.
We call these generalized affine loops because the $i$th iteration $\phi_{\text{\cinline{v}}_i}$ is of the form $ci + b$, where $b$ is the base case given by the initial value of the variable before the loop.
If $c$ is strictly increasing, \faultless also infers that $c \ge b$, and if $c$ is strictly decreasing, then \faultless infers that $c \le b$.

The necessary conditions for a change to a variable to be a generalized affine update are fairly strict, but the majority of modifications to stack variables in loops are generalized affine: on \realtype, 71.5\% of loop-$\phi$ variables meet the criteria and at least one invariant is inferred.

To infer these invariants, \faultless performs an additional execution of the code before the main execution.
The base cases and changes made during the loop are logged and used to compute the invariants.
Then during the main execution, the invariants are injected into the path condition.
If there are no loop-$\phi$ instructions in a function, the pre-execution is skipped.

\subsubsection{The Heap Replay Buffer}
\label{sec:heap_replay_buffer}

When a linked data structure is traversed in a loop, a loop-$\phi$ instruction captures the changing value of the pointer to the current part of the data structure during iteration.
Intuitively, these modifications should impact the initial pointer as well; the subsequent read of \cinline{head->val} should logically return 0 in Figure~\ref{fig:linked_list_init}.
But because the stack slot for the modified variable (\cinline{current}) is initialized with a mediloop $\phi$ symbolic variable in the loop's head, the modifications are made relative to the $\phi$ symbolic variable, not the value in \cinline{head} (which here would be $p_0$, the value of the first parameter).
So by default, the memory model would return a fresh derived variable for the \cinline{head->val} memory read.

\begin{figure}
\begin{subfigure}[t]{0.53\textwidth}
\begin{lstlisting}[style=cstyle,basicstyle=\ttfamily\bfseries\scriptsize]
struct node { int val; struct node * next; };
void foo(struct node * head) {
    struct node * current = head;
    while (current) {
      current->val = 0;
      current = current->next;
    }
    // ...
    int value = head->val;
    // ...
}
\end{lstlisting}
\caption{Initializing a linked list in C.}
\label{fig:linked_list_init:c}
\end{subfigure}
\begin{subfigure}[t]{0.46\textwidth}
\begin{lstlisting}[style=cstyle,basicstyle=\ttfamily\bfseries\scriptsize]
%0 = head
%1 = phi(%0, %5)
loop %1
    %3 = %1->val
    %4 = store %3 0
    %5 = %1->next
// ...
%2 = head->val
// ...
\end{lstlisting}
\caption{Figure~\ref{fig:linked_list_init:c} in \faultless IR. SSA variables are assigned names in the format \%X, following the style of LLVM IR.}
\label{fig:linked_list_init:ir}
\end{subfigure}
\caption{A function which initializes a linked list, then access a value written later from the head pointer.}
\label{fig:linked_list_init}
\end{figure}

To handle this situation, \faultless applies a \emph{replay buffer}.
We define the replay buffer as a function, $r(s)$, where $s$ is a symbolic variable.
The replay buffer captures reads and writes made to the heap relative to the mediloop $\phi$ symbolic variable ($r(\phi)$) through the duration of the loop.

At the beginning of the loop, the $\phi$ instruction records the base address $b$ of the $\phi$ instruction's base case as corresponding to that variable's mediloop $\phi$ symbolic variable.
In Figure~\ref{fig:linked_list_init}, it records that $\phi_{\text{\cinline{current}}}$ maps to $p_0$.
Then if the base case's base address is read from, the replay buffer first applies replayed writes to the heap, writing $r(b)$, or for Figure~\ref{fig:linked_list_init} specifically, $r(p_0)$, which is just \cinline{head->val = 0}.
\faultless uses reads in the buffer to apply the replay buffer recursively.
Thus \cinline{head->next->next->val} would also return 0.

\subsection{The Proof Engine}
\label{sec:proof_engine}

The proof engine uses information recorded during the execution to attempt to show that the necessary equivalence conditions (Section~\ref{sec:defining_equivalence}) hold.
We refer to the two functions being compared as the \emph{left} and \emph{right} functions.

In checking that the necessary equivalence conditions are met, \faultless often checks if two symbolic expressions, one from the left function and the other from the right (e.g. return values or callee arguments), are equivalent to each other.
Any expression can be viewed as a function of its \emph{free variables}; $x + y + 2$ is a function of $x$ and $y$, for instance.
It only makes sense to consider the equivalence of two expressions if they are defined in terms of the \emph{same} free variables.
Unfortunately, this is not the case for many pairs of expressions in \faultless because loop-$\phi$ instructions and function calls introduce fresh progenitor variables.
Thus, a prerequisite to checking the necessary equivalence conditions is building a \emph{coupling relation}, denoted $C$, between the two programs, populating it with pairs of symbolic variables shown to be equivalent to each other.
These pairs of variables can be considered ``the same'' for the purposes of determining if two expressions are equivalent.

The process of building a coupling relation often itself requires a (smaller) coupling relation.
For instance, in order to show two function calls equivalent (and thus add their progenitor variables to the coupling relation), it is necessary to show that each pair of arguments is positionally equivalent.
Those arguments are symbolic expressions themselves defined in terms of symbolic variables, whose coupling relationship must be known prior to proving equivalence.
Thus, building a coupling relation is an incremental affair that goes hand-in-hand with showing the necessary equivalence conditions.

\begin{figure}
\begin{subfigure}{0.49\textwidth}
\begin{lstlisting}[style=cstyle,basicstyle=\ttfamily\bfseries\scriptsize]
void scale_cell(double c[3][3], double s)
{
    if (s < 0)
        s = pow(-s/cell_volume(c), 1.0/3);
    for(int i = 0; i<3; ++i) {
        real_prod(c[i], s);
    }
}
\end{lstlisting}
\caption{A small function featuring several function calls and a loop. \cinline{cell_volume} is a dataflow dependency of \cinline{pow}, which in turn is a dataflow dependency of \cinline{read_prod}. The loop-$\phi$ for \cinline{i} is a dataflow and control dependency of \cinline{read_prod}.}
\label{fig:synchronization_graph:code}
\end{subfigure}
\hfill
\begin{subfigure}{0.49\textwidth}
\includegraphics[width=\textwidth]{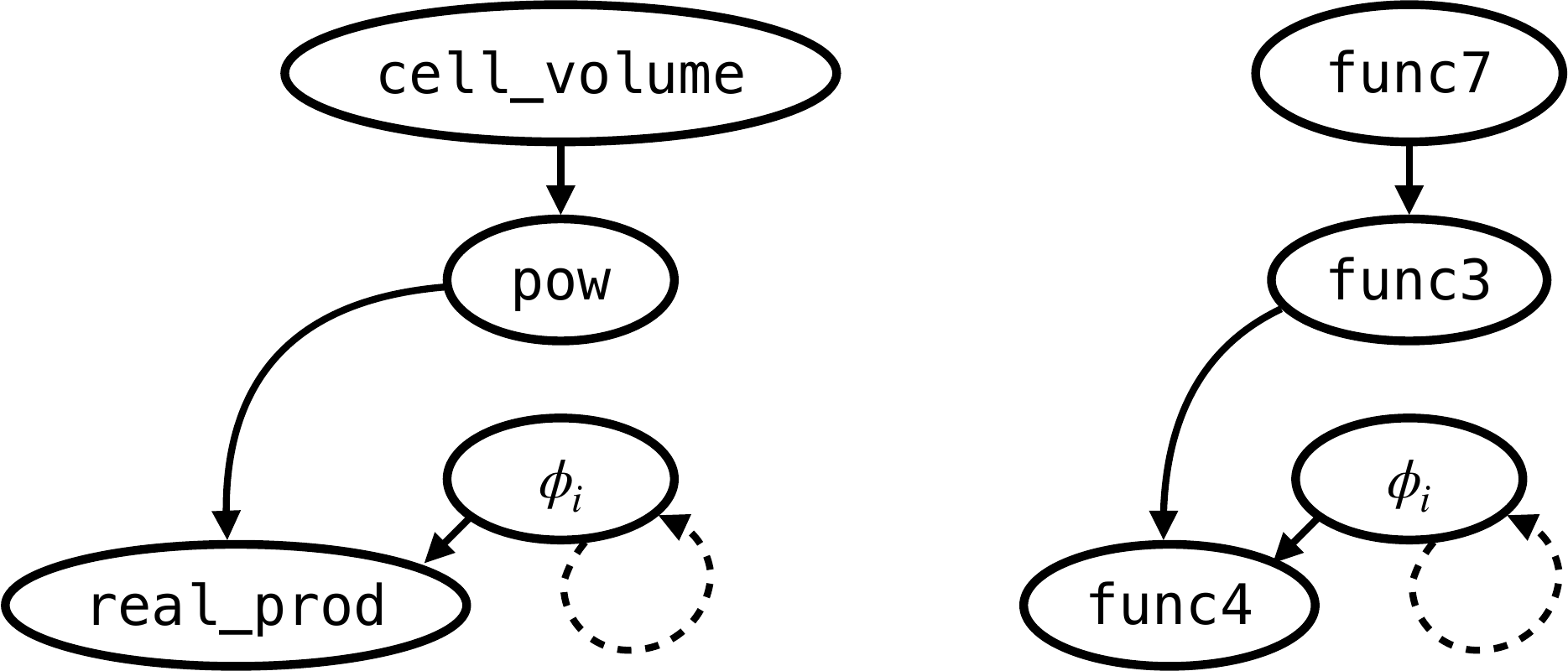}
\caption{The synchronization graphs for Figure~\ref{fig:synchronization_graph:code} (left), and a hypothetical decompilation of it (right). Dashed lines indicate a back-dependency edge going from the loop body to its head.}
\label{fig:synchronization_graphs:graphs}
\end{subfigure}
\caption{A function and the synchronization graph derived from it. A synchronization graph captures the control and dataflow dependencies between the instructions which introduce progenitor symbolic variables (Section~\ref{sec:derived_variables}), which we refer to as synchronization instructions.}
\label{fig:synchronization_graphs}
\end{figure}

\subsubsection{Synchronization Graphs}

The key data structure which structures the equivalence checking process is the \emph{synchronization graph}.
Figure~\ref{fig:synchronization_graphs} shows an example.
It is a directed, possibly cyclic, graph.
The left and right functions each have their own synchronization graphs.
The nodes in the graph are instructions that introduce progenitor symbolic variables (Section~\ref{sec:derived_variables}); that is, loop-$\phi$ instructions and function calls.
The edges represent transitive dataflow and/or control dependencies between the nodes: if there exists some combination of dataflow and/or control flow edges between two progenitor-instructions in the IR, then there is an edge between them in the synchronization graph.
If there is an edge from node $n$ to node $m$ in the synchronization graph, we say that $m$ is dependent on $n$, and that $n$ is a dependency of $m$.

Control dependencies are basic-block-level abstractions while synchronization graph nodes and dataflow dependencies are instruction-level abstractions.
To convert control-dependencies to instruction level-dependencies, \faultless adds an edge in the joint dataflow/control flow graph used to build the synchronization graph between each instruction in the dependent block and the branch instruction that induced the control flow.

The synchronization graph precisely captures the free variables in the expressions associated with each progenitor variable.
In Figure~\ref{fig:synchronization_graphs}, the execution of \cinline{cell_volume} produce a unique symbolic progenitor variable, $c_{\text{\cinline{cell_volume}}}$.
The symbolic expression for the first argument of \cinline{pow} is $-p_1 / c_{\text{\cinline{cell_volume}}}$, is defined in terms of this variable.
This is reflected in the edge between \cinline{cell_volume} and \cinline{pow} in the synchronization graph.

\subsubsection{Building the Coupling Relation}
\label{sec:building_the_coupling_relation}

\faultless builds the coupling relation from the left and right synchronization graphs $G_l$ and $G_r$.
An entry in the coupling relation consists of a pair of synchronization graph nodes---an element of the product graph---which have been shown to be equivalent to each other.
Only entries of the same operation type can be equivalent: a function call cannot be equivalent to a loop-$\phi$ instruction.
Then the coupling relation $C$ is a subset of the cartesian products of the call nodes and the $\phi$ nodes in the synchronization graphs:
\begin{equation}
\label{eq:coupling_relation_bound}
C \subseteq \{ (n_l, n_r)\; | \;type(n_l) = type(n_r) \wedge n_l \in G_l \wedge n_r \in G_r \}
\end{equation}
Na\"ively, \faultless would have to check each element of the product graph and determine if the pair of nodes is actually equivalent based on the logged execution data.
But enumerating all combinations is expensive, and as discussed previously, performing such equivalence checking often requires a smaller partial coupling relation over the free variables for the expressions involved.

Instead, \faultless uses the edges in $G_l$ and $G_r$ to structure the building process.
The coupling relation is initially empty.
The algorithm begins with \emph{independent nodes}: those associated with no free variables or whose free variables consist of only parameters.
In Figure~\ref{fig:synchronization_graphs}, the \cinline{cell_volume} and \cinline{func7} nodes are \emph{independent nodes}.
(By \faultless' induction system, discussed below, the $\phi_i$ instructions are actually independent nodes as well).
For each pair $(n_l, n_r)$ of same-operation, independent nodes, \faultless proposes the conjecture that the left and right elements of the pair are equivalent, e.g. \cinline{cell_volume} $=$ \cinline{func7}.
If a conjecture can be proven, \faultless adds the corresponding progenitor variables to the coupling relation and adds each pair of same-operation direct dependent nodes of $n_l$ and $n_r$ as conjectures if no such conjecture already exists.
In Figure~\ref{fig:synchronization_graphs}, if \cinline{cell_volume} is proven equivalent to \cinline{func7}, then $(c_{\text{\cinline{cell_volume}}}, c_{\text{\cinline{func7}}})$ will be added to the coupling relation and $(c_\text{\cinline{pow}}, c_\text{\cinline{func3}})$ will be added as a conjecture.
When each free variable in the left node of a conjecture is related to one in the right node and vice versa, \faultless will attempt to prove the conjecture.
If it can't but additional relational information between the nodes' free variables is later proven, then \faultless will attempt to re-prove that conjecture.
\faultless uses a worklist algorithm similar to the one used in codealign~\cite{codealign} to perform this process.

\vspace{4pt}
\noindent
\emph{Proving a Conjecture}.
To prove a conjecture that two function callees are equivalent to each other, \faultless checks the conditions defined in the necessary equivalence conditions (Section~\ref{sec:defining_equivalence}) under the coupling relation.
This includes that checking that each pair of arguments to the function calls is positionally equivalent, that heapspace memory reachable from each addressable pair of arguments is equivalent, and that the calls occur under the same control-flow conditions.
For the control flow check, \faultless uses only the components of the path condition that are transitive control dependencies of the function call because only those components actually influence the control flow conditions under which a given instruction executes.

Loop-$\phi$ instructions are not described in the necessary equivalence conditions because they are internal features of a function that do not, by themselves, produce externally observable behavior.
To prove two loop-$\phi$ nodes equivalent, the ``base case'' arguments from outside the loop must be equivalent, the ``recursive case'' arguments form inside the loop must be equivalent, and the path conditions required to reach any back edge from the start of the loop must be equivalent.

\vspace{4pt}
\noindent
\emph{Dependency Cycles and Induction}.
Loops can create cycles of dependencies in the synchronization graph.
Self-loops are common: in a basic \cinline{for (int i = 0; i < n; ++i)} loop like the one in Figure~\ref{fig:synchronization_graphs}, the recursive case for the $\phi_{\text{\cinline{i}}}$ node is $\phi_{\text{\cinline{i}}} + 1$; thus, there is a dataflow dependency between the $\phi_{\text{\cinline{i}}}$ instruction and itself.
In general, all recursive case arguments of a $\phi$ instruction must involve some kind of cycle.
Cycles mean the coupling relation can't contain the necessary variable relationships when attempting to prove the recursive arguments equivalent.
To solve this, \faultless uses induction to break cycles of dependencies, also mirroring codealign~\cite{codealign}.
For the base case of the induction, \faultless will attempt to show that the base case arguments of the two $\phi$ nodes are equivalent.
If it can, it makes the inductive hypothesis---if the values are equivalent at the start of the loop iteration, then the values at the end of the loop iteration are equivalent.
Concretely, \faultless provisionally adds an equivalence relationship between the $\phi$ variables to the coupling relation and then re-visits the node to prove the  recursive cases and path conditions equivalent with the updated coupling relation when it has finished proving relationships inside the loop.
If the inductive hypothesis cannot be proven, \faultless revokes the inductive assumption and anything from inside the loop proven based on the unprovable assumption.
For instance, if the $\phi_{\text{\cinline{i}}}$ instructions in the original and decompiled code in Figure~\ref{fig:synchronization_graphs} cannot be proven equivalent, then that assumption is revoked and the lemma \cinline{real_prod} == \cinline{func4} is also revoked if it exists.
To maintain soundness, \faultless only allows elements of the coupling relation that are transitive dependencies of a given conjecture for use in the proof of that conjecture; that way, if one of those dependencies was a bad inductive assumption, \faultless knows to revoke it.

\subsubsection{Heaplet equivalence}

Several of the necessary equivalence conditions require that memory reachable from a certain address e.g. (function arguments, the return value) is equivalent.
Determining whether or not this is the case is the job of the heaplet equivalence checker.
In other words, the heaplet equivalence checker implements support for McCarthy's extensionality axiom for arrays.
The heaplet equivalence checker takes as input a base address from the left function and one from the right function, as well as the heaps for each of these functions and a coupling relation.

Under the extensionality axiom, two arrays---or in our case, Address Spaces---are equivalent if they are equivalent at all indices.
To accomplish this, we read a value of the type of the base address, $t$, at an \emph{arbitrary} index $\epsilon$.
Concretely, we run the query $\text{heap.read}(\text{sizeof}(t) \cdot \epsilon, t)$.
Note that this implements the extensionality axiom with respect to a type as opposed to treating every memory cell as a collection of untyped bits.
Because \faultless offers the ability to represent integers as ``math'' integers rather than bitvectors, (see Table~\ref{tab:equivalence_options}) and floats as real numbers rather than IEEE-754 floating point numbers, this distinction is required to perform heaplet equivalence checking under all primitive type modes.
Math integers can't be represented as untyped bits.

The reads are done for both the left and right base addresses on their respective heaps, and the result is checked for equivalence subject to the coupling relation.

\vspace{4pt}
\noindent
\emph{Memory Formatting for Decompiled Code}.
One of the key challenges with decompiled code is the lack of user-defined types, especially \cinline{struct}s.
In the validation task, it is often the case that the base address is a pointer to a structure while in the decompiled code it is a pointer to a primitive type, usually a size-based placeholder type (see Section~\ref{sec:type_deduction}).
The resulting values read will be trivially nonequivalent because one is a composite data type and the other is a primitive data type.
To prevent this, \faultless provides the option to allow the decompiled code to be read with the original code's composite type.

\subsubsection{Global Assumptions}

Global variables represent inputs and outputs for a function.
Unlike parameters, however, which we can reliably relate to each other positionally on the parameter list, or return values, which are trivial to relate because a function can only return one value, it is not immediately clear how global variables relate to each other across the left and right functions, especially when both the left and right functions have multiple global variables.
And unlike $\phi$ or callee return values, globals cannot be aligned by the way that they are computed from function inputs, because globals themselves are  function inputs.

Instead, \faultless determines that two globals are the same if they are used in the same context.
When two symbolic expressions are checked for equivalence in the construction of the coupling relation and the expressions are nonequivalent, \faultless checks to see if there is one global variable involved in each of the expressions.
If so, then \faultless checks to see if equating those two global variables results in the expression being equivalent.
If it does, then \faultless assumes that those variables are equivalent and records this assumption.

While it's difficult to tell which global variables correspond to each other, it is sometimes easy to tell which globals do \emph{not} correspond to each other by looking at the values stored in the globals when the function terminates: globals holding different values are definitely different variables.
\faultless rejects any assumed equivalence relationship between globals that store different values.

\subsubsection{Checking Equivalence Conditions}
\label{sec:checking_equivalence_conditions}

In building the coupling relation, \faultless has already started to satisfy the necessary equivalence conditions (Section~\ref{sec:defining_equivalence})  by showing certain function calls in the left function equivalent to those in the right function.
But there are other necessary equivalence conditions that must be satisfied to declare the left and right function equivalent.

For callees, \faultless must show that \emph{each} function call in the left function is equivalent to at least one in the right function and vice versa.

Using the entire coupling relation, \faultless attempts to prove the return values equivalent.
If both return values are addressable, it attempts to prove that the heaplets reachable from the left and right base address are equivalent.
If one is addressable and the other is not, \faultless returns that the left and right function are not equivalent.
\faultless also checks that memory reachable from each input parameter in the left function is equivalent to the memory reachable from the corresponding input parameter in the right function using the heaplet equivalence checker.

\vspace{4pt}
\noindent
\emph{Consistency Checking}.
Finally, \faultless must ensure that function and global variable names are consistent across the left and right function; that is, that the function and global variable names form a bijective mapping. Each of these name types is checked for consistency separately.

To do this, \faultless sets up a constraint satisfaction problem.
Each name is assigned a unique integer variable: $l_i$ for the left function and $r_i$ for the right function.
Within a given function, distinct names are constrained to have different integer variables: $l_i \ne l_j$.
The cross-functional constraints are somewhat more complicated.
Let $l_i$ be the name variable of a call in the left function.
Then let $R$ be the set of name variables in the right function for calls that have been shown equivalent in the coupling relation (for function names) or assumed equivalent (for global variable names).
Then \faultless adds the constraint $\bigvee_{r \in R} l_i = r$.
\faultless also adds its mirror image on the right.

The generalized-disjunctive form of the cross-functional constraints is important to handle the case when there are multiple possible bijective mappings amongst the names.
For instance, consider decompiled code containing the calls \cinline{func1(0)} and \cinline{func2(0)} which a neural decompiler might rename to \cinline{foo(0)} and \cinline{bar(0)}.
Here, by arguments, these are all equivalent.
There are multiple possible bijective mappings: $\text{\cinline{func1}} \leftrightarrow \text{\cinline{foo}}$ and $\text{\cinline{func2}} \leftrightarrow \text{\cinline{bar}}$ or $\text{\cinline{func1}} \leftrightarrow \text{\cinline{bar}}$ and $\text{\cinline{func2}} \leftrightarrow \text{\cinline{foo}}$.
If we forced every name variable in the left function to be equivalent to the name variable on the right, the constraint problem would be unsatisfiable:
\begin{equation*}
l_{\text{\cinline{func1}}} = r_{\text{\cinline{foo}}} \wedge l_{\text{\cinline{func2}}} = r_{\text{\cinline{foo}}} \wedge l_{\text{\cinline{func1}}} \ne l_{\text{\cinline{func2}}}.
\end{equation*}
The disjunctive constraint prevents this issue.

\section{Evaluation}
\label{sec:evaluation}

We ask three research questions to empirically characterize \faultless' capabilities.

Because program equivalence is undecidable, any program equivalence technique must be sound, complete, or neither, but not both.
Faultless is designed to be sound to foster the reverse engineer's trust of the neural decompiler by ensuring that there are no semantic mistakes in the neural decompiler's prediction.
A consequence of this design decision is that it is necessarily incomplete: there are equivalent pairs of code which are the technique cannot prove equivalent.
Therefore, we ask \emph{\textbf{RQ1:} How complete is \faultless?}
That is, how often can \faultless prove equivalent pairs of functions equivalent?

There are some known soundness limitations in its current implementation (see Section~\ref{sec:limitations:soundness}).
Therefore, we also ask \emph{\textbf{RQ2:} How frequently do \faultless' soundness limitations cause and unsound equivalence judgment?}

Finally, on the validation task, \faultless is intended to be used live by reverse engineers as part of their tool suite.
That means runtime performance is especially important, so we ask 
\emph{\textbf{RQ3:} What are the runtime performance characteristics of \faultless?}

We answer these questions with data from two experiments using two complementary oracles.
We observe that the original code and deterministically decompiled code for the same function are equivalent by construction (excepting some decompiler unsoundness), but significantly syntactically different.
In the first experiment, we use this ``deterministic transformation'' oracle.
However, because the pairs of functions are all equivalent, this doesn't provide any pairs of nonequivalent functions for answering RQ2.
Instead, in the second experiment, we use unit tests as an oracle.
They're unsound (there could be some differentiating behavior between the two functions that the tests do not exercise) but complete.
That means that if the unit tests return that two functions are \emph{not} equivalent, they aren't; the test case serves as a counterexample.

\subsection{Experimental Design}

We conduct two experiments, one using each of the two complementary oracles.
In both experiments, we use codealign's~\cite{codealign} dependency-based equivalence as a baseline.
We are not aware of other techniques designed to check the equivalence of C functions in isolation without the rest of the codebase, especially for decompiled and predicted code---a key motivator for the existence of \faultless in the first place.

\subsubsection{Experiment 1: Deterministic Transformation Oracle}
\label{sec:deterministic_transformation_oracle}

The deterministic transformation oracle uses compiler and decompiler transformations to provide equivalent pairs of code.
Dramko et al.~\cite{idioms} introduce \realtype, which already contains these pairs, there as labeled training data for a machine learning model.
We reused \realtype for this experiment, though we rebuilt it for these experiments to include definitions of types referenced in the type descriptors of the function that are not part of a variable declaration.
(\idioms only predicted variables' types so those type definitions were unnecessary.)

We use the test set in the experiments.
We filter from the test set examples that include unsupported features, including \cinline{goto}s, \cinline{union}s, \cinline{enum}s, the variadic parameter API, inline assembly, compiler extensions, SIMD vector types and \cinline{#pragma} directives that were not handled in the preprocessor run when constructing the dataset.
\faultless does not currently support Hex-Rays' partial-object macros (like \cinline{LODWORD}).
Also unsupported is the subtraction of two pointers, a rarely used feature of C, which is only defined if those pointers point to the same array object; under context independence, it usually cannot be determined if two pointers do actually point to the same object.
When an array type is declared in a header and its length is defined not as integer but as an arbitrary C expression, the \realtype dataset construction script provides a sentinel array length of $-1$.
This is of course an invalid length so \faultless filters these examples out.
Finally, sometimes the Hex-Rays decompiler will recognize and define variables of standard-library types like \cinline{jmp_buf} and \cinline{__compar_fn_t}.
We provided a header of many standard library types (all those that show up in the decompiled code in the validation set), but there are some types in the test set that aren't in the validation set.
After this filtering, the rebuilt test set's 2229 examples were reduced to 1565.

In the experiment, we run \faultless on each original/decompiled pair and report whether \faultless reports the functions equivalent, nonequivalent, or crashes.
We also record the time it takes to perform the equivalence checking.
We provide the definitions of UDTs used in the functions along with the original code.

We also perform some normalization modifications prior to calling \faultless on each example pair.
The compiler or decompiler sometimes replaces \cinline{printf} with \cinline{puts}, \cinline{putchar}, \cinline{fputc}, or \cinline{fwrite}, adjusting the arguments to match the function.
When these functions do not appear in the original code but do appear in the decompiled code, we normalize the functions in the decompiled code back to \cinline{printf}.
Additionally, we normalize decompiler fidelity issue C4.c. from Dramko et al.'s taxonomy of C decompiler fidelity issues~\cite{taxonomy}, which refers to the pattern where the decompiled code includes an extra address-of \cinline{&} around global variables in certain circumstances, like function calls and return values.
If is is observed in the decompiled code, the \cinline{&} is stripped out.

In general, we try to use math-int ($\mathbb{Z}$) representations of integers rather than bitvectors, if possible.
Math-int modeling allows \faultless to abstract away issues with imprecise primitive type recovery, putting the emphasis on showing the algorithms equivalent.
However, bitwise operations like \cinline{>>} and \cinline{|} are not defined on math integers, so we switch to bitvectors if any such operations occur in either function.

Deterministic decompilers are known to be unsound.
Many instances of unsoundness follow systematic patterns, which \faultless is designed to handle via the options in Table~\ref{tab:equivalence_options}.
However, some instances of unsoundness do not fit any such pattern; the decompiled code is truly just incorrect.
This somewhat undermines the reliability of the deterministic transformation oracle.
To account for this, we manually analyzed the results for a random sample of the predictions.

\subsubsection{Experiment 2: Unit Test Oracle}

For the unit test oracle, we use \exebench~\cite{exebench}, a dataset of C functions mined from GitHub.
Its test and validation sets have unit tests bundled with each example.
\exebench provides, then, the unit test oracle itself as well as the original code.
For the functions to compare with the original code, we use \idioms' predictions (particularly those from the best performing model, codegemma-7b).
Code predicted by a neural decompiler is exactly the sort of thing that \faultless is designed to handle.

\exebench is inconsistent with itself on some examples; the original code functions don't always pass the unit tests included in the dataset.
We filter out these examples.
We also filter out examples with unsupported features as in Experiment 1.
After filtering, there are 881 examples remaining of 1741.

In this experiment, we turn on the option which requires two functions to have equivalent names (instead of just consistent names) to be considered equivalent.
This better reflects the behavior of the unit test oracle, which counts differently-named functions as incorrect.

% Evaluation results

\begin{table}
\centering
\caption{Experiment 1 Results: Completeness}
\label{tab:completeness}
{\small
\begin{tabular}{llrrrrrr}
\toprule
 &  & \multicolumn{2}{c}{Correct} & \multicolumn{2}{c}{Incorrect} & \multicolumn{2}{c}{Crash} \\
\midrule
\multirow{3}{*}{O0} & \faultless & 805 & (51.4\%) & 495 & (31.6\%) & 265 & (16.9\%) \\
& Dependency-Based Equivalence & 69 & (4.4\%) & 1273 & (81.3\%) & 223 & (14.2\%) \\
& Dependency-Based Equiv. (Consistency)& 109 & (7.0\%) & 1233 & (78.8\%) & 223 & (14.2\%) \\
\midrule
\multirow{3}{*}{O1} & \faultless & 664 & (45.1\%) & 491 & (33.4\%) & 316 & (21.5\%) \\
& Dependency-Based Equivalence  & 65 & (4.4\%) & 1199 & (81.5\%) & 207 & (14.1\%) \\
& Dependency-Based Equiv. (Consistency)& 85 & (5.8\%) & 1179 & (80.1\%) & 207 & (14.1\%) \\
\midrule
\multirow{3}{*}{O2} & \faultless & 599 & (43.1\%) & 515 & (37.1\%) & 276 & (19.9\%) \\
& Dependency-Based Equivalence  & 69 & (5.0\%) & 1094 & (78.7\%) & 227 & (16.3\%) \\
& Dependency-Based Equiv. (Consistency) & 88 & (6.3\%) & 1075 & (77.3\%) & 227 & (16.3\%) \\
\midrule
\multirow{3}{*}{O3} & \faultless & 581 & (44.7\%) & 491 & (37.7\%) & 229 & (17.6\%) \\
& Dependency-Based Equivalence  & 70 & (5.4\%) & 1009 & (77.6\%) & 222 & (17.1\%) \\
& Dependency-Based Equiv. (Consistency) & 86 & (6.6\%) & 993 & (76.3\%) & 222 & (17.1\%) \\
\bottomrule
\end{tabular}
}

\caption*{Results of the completeness experiment.
All pairs of examples are equivalent by construction; scores reflect how many pairs each technique was able to prove equivalent (correct), unable to prove equivalent (incorrect), or which crashed on the input (crash).
Codealign~\cite{codealign} is used for dependency-based equivalence.}
\end{table}

\begin{table}
\caption{Experiment 2 Results: Comparison with Unit Tests}
\label{tab:soundness}
\begin{subtable}{0.49\textwidth}
\caption{\faultless}
\label{tab:soundness:faultless}
{\small
\begin{tabular}{|l|r|r|r|}
\toprule
     & equivalent & not equivalent & crash \\
     \midrule
tests pass  & 525 &  81 & 168 \\
\hline
tests fail & 6 & 70 & 31 \\
\bottomrule
\end{tabular}
} % end \small
\vspace{2pt}

\centering
Agreement: 87.243\%
\end{subtable}
\begin{subtable}{0.49\textwidth}
\caption{Dependency-Based Equivalence}
{\small
\begin{tabular}{|l|r|r|r|}
\toprule
          &  equivalent & not equivalent & crash \\
\midrule
tests pass & 367 & 402 & 5 \\
\hline
tests fail & 1 & 105 & 1 \\
\bottomrule
\end{tabular}
} % end \small
\vspace{2pt}

\centering
Agreement: 53.942\%
\end{subtable}
\caption*{Results of Experiment 2, comparing \faultless and codealign with unit tests, including confusion matrices and agreement scores. The agreement scores are the sum of diagonal, where both unit tests and the equivalence technique agree that the code is equivalent or not, divided by the sum of all four noncrash cells of the matrix. Function-name consistency is disabled to better conform to the unit test oracle. The bottom left cell represents instances of unsoundness.}
\end{table}

\subsection{Results}

The results are shown in Table~\ref{tab:completeness} for experiment 1 and Table~\ref{tab:soundness} for experiment 2.
The runtime performance results are in Figure~\ref{fig:runtimes}.

\subsubsection{RQ1: How complete is \faultless?}

\faultless is able to prove equivalent above 50\% of the equivalent original/decompiled code pairs equivalent at O0, and around 43--45\% at the other optimization levels as shown in Table~\ref{tab:completeness}.
\faultless greatly outperforms codealign~\cite{codealign} in this experiment, which simulates performing the validation task on a perfect neural decompiler, because dependency-based equivalence is too brittle to handle features of decompiled code like desugared field-access operations.
On \exebench, (Table~\ref{tab:soundness}) \faultless agrees with unit tests 87.2\% of the time, substantially closing the completeness gap between sound codealign and unsound unit tests, which are an overapproximation of equivalence.

\faultless has a higher success rate in Experiment 2 on \exebench than in Experiment 1 on \realtype because \realtype is a more challenging dataset containing the full spectrum of C complexity.
We sampled a set of 30 original/decompiled function pairs judged nonequivalent by \faultless and manually analyzed the reasons for failures.
Many of these---17, over half---are fairly mundane and unrelated to \faultless' novel modeling and equivalence checking contributions.
Five are \faultless compiler bugs, three are \cinline{printf} canonicalization bugs, seven are dataset generation issues, and two contain static variables, which \faultless doesn't currently support.
This suggests that with additional implementation effort, \faultless has room to improve its performance.

Four of the 30 cases were from oracle failures, where the decompiler is \emph{not} semantics preserving, even when accounting for the systematic issues identified in Dramko et al.'s taxonomy~\cite{taxonomy} with the options in Table~\ref{tab:equivalence_options}.
\faultless correctly rejects these as not equivalent, whereas the experiment presumes them to be equivalent.
While this was not the motivation behind the design of \faultless, it is fairly effective as a decompiler-testing technique like D-Helix~\cite{zou2024d}, helping to locate bugs in decompiled code.

The final nine cases are fundamental completeness limitations in \faultless.
\faultless must necessarily be incomplete.
Three of these examples highlight the limitations of operating in a context-independent environment---while the decompiled code is not technically incorrect, \faultless does not have enough information to soundly show the functions equivalent.
\faultless' completeness limitations are discussed in detail in Section~\ref{sec:limitations:completeness}.

The crash rate of around 17\% reported in Table~\ref{tab:completeness} is relatively high.
A minority of these crashes, 36 in total, are due to limitations in the addressability model: attempting to access an address for which the base address cannot be uniquely determined.
The majority of these cases occur when there are base addresses from different paths encapsulated in an if-then-else expression.
The fix is conceptually simple: access both base addresses subject to the additional read/write condition of the if-condition (for the then branch) or its negation (for the else branch).
Other crashes are due to corner cases in compiling and modeling the full breadth of the complex C programming language.

Codealign is designed for the evaluation task while this experiment simulates the validation task on perfect predictions; as a result, its crash rate is around five times higher than reported in the original paper due to difficulties with processing decompiled code.

\begin{tcolorbox}[width=\columnwidth, boxsep=0pt, left=4pt, right=7pt, top=7pt, arc=4pt, boxrule=1pt, toprule=1pt, colback=white]%%
\noindent
Answer to \textbf{RQ1}: Faultless is able to prove equivalent more than 50\% of equivalent pairs of decompiled/original functions in unoptimized code and more than 43\% in optimized code.
\end{tcolorbox}

\subsubsection{RQ2: How frequently do \faultless' soundness limitations cause and unsound equivalence judgment?}

Table~\ref{tab:soundness:faultless} shows that \faultless' soundness limitations are relatively uncommon, at least on \exebench.
Virtually all of the disagreement is in the ``tests pass''/``\faultless reports nonequivalent'' category on the top right of the confusion matrix, which represents incompleteness.
Of the six instances of unsoundness in the bottom left corner of the confusion matrix only five are real: one is due to undefined behavior in the \exebench-provided harness and test case; the actual generated functions are equivalent.
We discuss soundness limitations further in Section~\ref{sec:limitations:soundness}; all of the instances of unsoundness fall into categories discussed there.

\begin{tcolorbox}[width=\columnwidth, boxsep=0pt, left=4pt, right=7pt, top=7pt, arc=4pt, boxrule=1pt, toprule=1pt, colback=white]%%
\noindent
Answer to \textbf{RQ2}: \faultless' soundness limitations are exploited less than 1\% of the time on \exebench.
\end{tcolorbox}

\begin{figure}
\begin{subfigure}{0.49\textwidth}
\caption{\faultless Runtimes}
\label{fig:noncrash_runtimes}
\includegraphics[width=\textwidth]{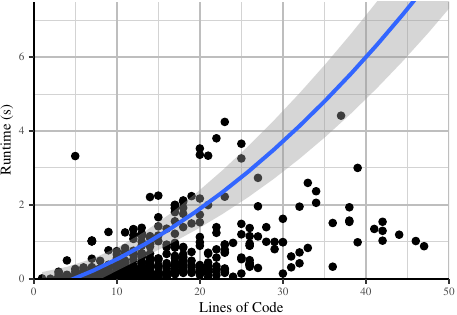}
\end{subfigure}
\hfill
\begin{subfigure}{0.49\textwidth}
\caption{\faultless Runtimes: Equivalent Only}
\label{fig:success_runtimes}
\includegraphics[width=\textwidth]{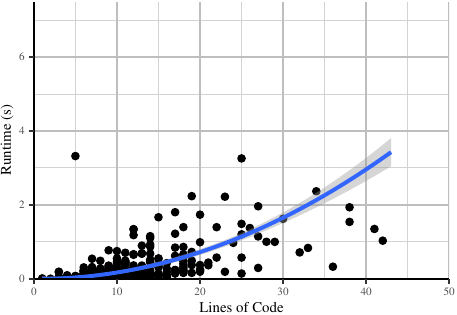}
\end{subfigure}
\caption{Runtime efficiency for \faultless on Experiment 1 with quadratic regression trend lines. Crash runtimes are excluded because crashes usually happen before the expensive proof engine stage. Figure~\ref{fig:noncrash_runtimes} includes pairs of functions that \faultless reported nonequivalent. The trend line in Figure~\ref{fig:noncrash_runtimes} is pulled up by a few outliers with runtimes greater than 7.5 seconds.}
\label{fig:runtimes}
\end{figure}

\subsubsection{RQ3: What are the runtime performance characteristics of \faultless?}

As shown in Figure~\ref{fig:runtimes}, \faultless is generally very fast.
\faultless can produce an equivalence judgement in under one second for 90.1\% of examples in \realtype and in under two seconds for 96.7\% of examples.

SMT solver queries dominate the runtime of \faultless, so it follows that functions which produce more or larger solver queries constitute the remaining 3.3\% of high runtimes.
There are two main culprits.
The first is large synchronization graphs with many synchronization nodes---function calls or loop-$\phi$ instructions---with equivalent dependencies.
If there are $n$ calls in each function and each has no dependencies, the proof engine will perform $n^2$ call equivalence checks, which is expensive.
In contrast, if the $i$th call was a dependency of the $i + 1$st call, then the proof engine will have to perform only $n$ call equivalence checks.
The dependency structure can substantially influence \faultless' runtime.
Most dependency structures produce relatively small sets of synchronization nodes with the same dependencies, but exceptions exist.

The second is large fixed size arrays (e.g. stack allocated or as a struct field) in circumstances where the entire array needs to be read from memory, such as when a struct with a large fixed-size array member is passed by value to a callee.
Reads for full composite data types are decomposed into reads of each of their members.
That means $n$ reads for an array of length $n$.
Then suppose there are $m$ writes relative to the base address for that array.
That means that in the worst case, there are $2nm$ solver queries, an overlap query (Equation~\ref{eq:overlap}) and a coverage query (Equation~\ref{eq:coverage}) for each Write node, plus any additional queries for Join nodes.
This quickly becomes expensive.

\begin{tcolorbox}[width=\columnwidth, boxsep=0pt, left=4pt, right=7pt, top=7pt, arc=4pt, boxrule=1pt, toprule=1pt, colback=white]%%
\noindent
Answer to \textbf{RQ3}: \faultless is very fast, producing equivalence judgements in under one
second for 90.1\% of functions and in under two seconds for 96.7\% of functions on \realtype.
\end{tcolorbox}

\section{Utility}
\label{sec:utility}

\faultless' ability to compare decompiled, predicted, and original code allow it to fulfill multiple roles in both research and practice, assisting in both evaluation and validation tasks.
In this section, we show \faultless' utility on these tasks.

Recently, large language models have anecdotally become increasingly effective at performing neural decompilation.
But it has been difficult to measure how good they have become.
\faultless offers the ability to quickly and effectively benchmark these models' correctness.
In this section, we perform a series of experiments on a frontier model, GPT-5.5.
To begin, we ask \emph{\textbf{RQ4}: How effective is a modern frontier model on neural decompilation?}

Dramko et al.~\cite{idioms} showed that for smaller, fine-tuned or adapted models, adding neighboring context leads to an increase in model performance.
The motivation for adding neighboring context is independent of model size or architecture, though; it's a question of information: there is sometimes not enough information within a compiled function to correctly decompile that function, especially when it comes to user-defined types.
While \faultless does not evaluate UDT predictions directly, they are factored into the equivalence judgements by the offsets to which their fields map.
We use \faultless to help answer the research question \emph{\textbf{RQ5}: To what degree does neighboring context improve the correctness of a frontier model's predictions?}

One of the key challenges that \faultless faces is optimizations which transform loop structures, like loop unrolling or loop tiling.
By requiring $\phi$ instructions to be equivalent to prove values and calls affected by loops equivalent, \faultless effectively requires one iteration of one loop in the left function to correspond to one in the right function.
In traditional neural decompilation, the model is trained to predict the original code, thereby \emph{undoing} the optimizations.
This is problematic for validation because \faultless won't be able to prove even a perfectly correct prediction equivalent.
However, with large language models, one mitigation is to ask the model to preserve loop structure while otherwise making the code match the likely original source code as closely as possible, thereby offering the benefit of cleaned-up neurally decompiled code while also gaining the correctness assurances that \faultless provides.
We use \faultless to help answer the research question \emph{\textbf{RQ6}: To what degree does preserving loop structure increase \faultless' validation effectiveness?}

\faultless' validation equivalence judgements can be used by reverse engineers to reduce the risk of receiving incorrect code from a neural decompiler.
But these scores need not only be used by reverse engineers: \faultless' equivalence judgement and textual description of the reason for nonequivalence can also be used to provide feedback to the model to correct its prediction.
Thus, we ask the research question \emph{\textbf{RQ7}: How effective is \faultless at improving neural decompilers' predictions through re-prompting?}

\subsection{Experiment Methodology}

To perform the experiments, we use \realtype and prompt the frontier model GPT-5.5.
We use the same filtering as in Section~\ref{sec:deterministic_transformation_oracle}.
RQ6 requires performance data on optimized code; to save API costs, we chose one optimization level, O2, because it is the most difficult for \faultless in Table~\ref{tab:completeness}.
To answer the four research questions in this section, we perform seven different experiment runs with the settings:
\begin{itemize}
\item O0, function-level context
\item O0, neighbors-level context
\item O0, neighbors-level context with re-prompting
\item O2, function-level context
\item O2, neighbors-level context
\item O2, neighbors-level context with loop preservation prompting
\item O2, neighbors-level context with loop preservation prompting and re-prompting
\end{itemize}
For the loop-preservation prompting experiments, we append the following text to the prompt: 
\begin{quote}
Preserve the loop structure shown in the decompiled target. Do not attempt to undo loop-transforming compiler optimizations that may have changed the number, nesting, ordering, or boundaries of loops. Otherwise, make the code match the likely original source as closely as possible.
\end{quote}
For the re-prompting experiments, we re-prompt up to three times before giving up; re-prompting only occurs if \faultless judges the prediction to be incorrect.

In these experiments, we have all three source-level representations from Figure~\ref{fig:code_representations}.
Therefore, it's possible to judge the models on how well they would perform in validation, how well the prediction matches the decompiled code, and evaluation, how well it matches the original code.

When reporting scores, we exclude \faultless crashes.
In this section, we are not evaluating \faultless (as we were in Section~\ref{sec:evaluation}), rather, we are using \faultless to evaluate (and validate) a large language model.
\faultless crashes reflect bugs in or limitations of \faultless and do not provide any useful information on the correctness of the model's prediction.

\begin{table}
\caption{GPT-5-5 Neural Decompilation Performance}
\label{tab:gpt5.5}
\begin{subtable}{\textwidth}
\label{tab:utility:O0}
\caption{O0 Results}
\centering
\begin{tabular}{lrr}
\toprule
Context Type and Experiment Settings & Evaluation Accuracy & Validation Accuracy \\
\midrule
functions & 56.5\% & 78.6\% \\
neighbors & 55.9\% & 73.5\% \\ 
neighbors (reprompting) & 58.2\% & 92.1\% \\
\bottomrule
\end{tabular}
\end{subtable}

\vspace{4pt}
\begin{subtable}{\textwidth}
\caption{O2 Results}
\begin{tabular}{lrr}
\toprule
Context Type and Experiment Settings & Evaluation Accuracy & Validation Accuracy \\
\midrule
functions & 46.2\% & 74.1\% \\
neighbors & 48.7\% & 72.8\% \\
neighbors (loop preservation) & 48.3\% & 78.1\% \\
neighbors (loop preservation \& reprompting) & 48.0\% & 92.4\% \\
\bottomrule
\end{tabular}
\end{subtable}

\end{table}

\subsection{Results}

The experiment results are shown in Table~\ref{tab:gpt5.5}.

\subsubsection{RQ4: Evaluation of A Frontier Model}

The frontier model GPT-5.5 performs well at neural decompilation, scoring above 55\% accuracy on evaluation and 73\% on validation on unoptimized code and at least 46\% on evaluation and 72\% on validation at O2.
The fact that the validation scores are higher may be surprising, because neural decompilers are supposed to predict the original code and clean up decompiler artifacts in the decompiled code.
While GPT-5.5 does certainly clean up decompiler artifacts and the code looks reasonable, the structure of the code it generates more closely matches the decompiled code than the original code.
For instance, despite disabling inlining via command line arguments, some functions are still inlined in \realtype.
(In at least some instances, we observed this was due to an always-inline attribute present in the original code.)
To succeed in evaluation, the model would have to recognize which expressions or statements correspond to an inlined function, extract the function, and choose the correct argument order.
Meanwhile, to succeed at validation, the function need only propagate the body of the inlined function in the decompiled code to the predicted code like any other statement or expression---a much easier task!

The scores here are also an underestimate of GPT-5.5's performance because \faultless is incomplete.
While the model does make mistakes, the majority of \faultless' nonequivalence judetments we observed are due to its incompleteness; the predictions were actually correct.

\begin{tcolorbox}[width=\columnwidth, boxsep=0pt, left=4pt, right=7pt, top=7pt, arc=4pt, boxrule=1pt, toprule=1pt, colback=white]%%
\noindent
Answer to \textbf{RQ4}: Frontier models are very proficient at recovering semantics in neural decompilation, scoring above
70\% validation accuracy on \faultless.
\end{tcolorbox}

\subsubsection{RQ5: Function-Level vs. Neighboring Context for a Frontier Model}
\label{sec:rq5}

Central to \idioms~\cite{idioms} design and the driving force behind the addition of neighboring context is the scattered evidence problem for UDTs.
In this experiment, we see how neighboring context impacts the much larger GPT-5.5.

Surprisingly, the addition of neighboring context has little effect, or actually \emph{decreases} overall performance: validation accuracy decreases from 78.5\% to 73.5\% at O0 and 74.1\% to 72.8\% at O2.
Some functions are entirely self-contained and need no additional context, and irrelevant context has been shown to distract large language models~\cite{shi2023llmdistraction}.
The scattered evidence problem means there are fundamental limits to the UDT definitions that the LLM can recover, and incorrect offsets, will, if they are related to a function behavior that \faultless checks, result in nonequivalence.
However, we frequently observed the LLM, unbidden, inserting padding fields of the appropriate size to make the offsets correct, preventing this from being an issue.
Additionally, the LLM has memorized the definitions of all standard-library \cinline{struct}s, making it easy to bypass the scattered evidence problem in these instances and predict the entire \cinline{struct} correctly.

\begin{tcolorbox}[width=\columnwidth, boxsep=0pt, left=4pt, right=7pt, top=7pt, arc=4pt, boxrule=1pt, toprule=1pt, colback=white]%%
\noindent
Answer to \textbf{RQ5}: Neighboring context does not help improve frontier models’ \faultless scores, though it is necessary for full UDT definitions in some cases.
\end{tcolorbox}

\subsubsection{RQ6: The Impact of Loop Structure Preservation}
\label{sec:rq6}

A key limitation of \faultless is that it requires on loop in the left function to correspond to one in the right and vice versa.
Prompting the model to preserve loop structure helps to mitigate this limitation.

In particular, we see that adding the loop preservation prompt increases validation scores modestly from 72.1\% to 78.1\%.
The baseline scores are already fairly high.
Not all loops are impacted by loop optimizations and many functions are relatively small (the median length for \realtype is 14.2 lines of code) and contain no or one loop, so the probability of having a loop with unaffected by structure-altering optimizations is reasonably high.
Combined with GPT-5.5's propensity to conform to the decompiled code, this means that it can score relatively well without the additional prodding in the prompt.
However, for functions with loops that are impacted, structure preservation can help.

Evaluation scores are largely unaffected but decrease slightly; this is in line with expectations because the preservation of optimized loop structures makes the code less like the unoptimized original source code and therefore less likely to be judged equal to the original code.

\begin{tcolorbox}[width=\columnwidth, boxsep=0pt, left=4pt, right=7pt, top=7pt, arc=4pt, boxrule=1pt, toprule=1pt, colback=white]%%
\noindent
Answer to \textbf{RQ6}: Loop preservation prompting provides a modest increase in validation accuracy from an already high baseline, raising it from 72.1\% to 78.1\%.
\end{tcolorbox}

\subsubsection{RQ7: Reprompting Based on \faultless Feedback}
\label{sec:rq7}

\faultless can provide quick equivalence judgements along with error messages indicating the reason for nonequivalence (i.e. differing return values, a specific callee in one function doesn't match one in the other, etc.).
These properties make it suitable for re-prompting the LLM if \faultless is unable to prove the LLM's prediction correct.
Table~\ref{tab:gpt5.5} shows that re-prompting offers significant validation performance gains over the baselines, jumping from 73.5\% to 92.1\% at O0 and from 78.1\% to 92.4\% at O2.

\begin{figure}
\begin{subfigure}[t]{0.50\textwidth}
\begin{lstlisting}[style=cstyle,basicstyle=\ttfamily\bfseries\scriptsize]
void *hash_find(struct hash *h, void *item)
{
    int index = hash_find_index(h, item);
    
    if (index == -1)
        return ((void *)0);
        
    void *out = gap_get(h->data, index);
    
    return out;
}
\end{lstlisting}
\caption{The Original Code}
\label{fig:reprompting:original}
\end{subfigure}
\begin{subfigure}[t]{0.49\textwidth}
\begin{lstlisting}[style=cstyle,basicstyle=\ttfamily\bfseries\scriptsize]
__int64 func5(__int64 a1, __int64 a2) {
  __int64 result; // rax
  int v3;         // [rsp+1Ch] [rbp-4h]

  v3 = func4(a1, a2);
  if (v3 == -1)
    result = 0LL;
  else
    result = func1(*(_QWORD *)(a1+8), v3);
  return result;
}
\end{lstlisting}
\caption{The Decompiled Code}
\label{fig:reprompting:decompiled}
\end{subfigure}

\begin{subfigure}[t]{0.50\textwidth}
\begin{lstlisting}[style=cstyle,basicstyle=\ttfamily\bfseries\scriptsize]
void *hash_table_lookup(struct hash_table *table, void *key) {
  int index;

  index = hash_table_find_index(table, key);
  if (index == -1)
      return 0;

  return table->entries[index];
}
\end{lstlisting}
\caption{GPT 5.5's Initial Prediction}
\label{fig:reprompting:prediction1}
\end{subfigure}
\begin{subfigure}[t]{0.49\textwidth}
\begin{lstlisting}[style=cstyle,basicstyle=\ttfamily\bfseries\scriptsize]
void *hash_table_lookup(struct hash_table *table, void *key) {
  int index = hash_table_find_index(table, key);

  if (index == -1)
      return 0;

  return hash_table_get_entry(table->entries, index);
}
\end{lstlisting}
\caption{GPT 5.5's Corrected Prediction}
\label{fig:reprompting:prediction2}
\end{subfigure}
\caption{An example of a successful instance of re-prompting. GPT 5.5's initial prediction based on Figure~\ref{fig:reprompting:decompiled} is shown in Figure~\ref{fig:reprompting:prediction1}. This prediction is not correct; it's missing the call to \cinline{gap_get}. (\cinline{gap_get} is not simply \cinline{h->data[index]}; it performs additional checks.) \faultless reports that the call is missing, and when prompted again the call is added as shown in Figure~\ref{fig:reprompting:prediction2}. \faultless then judges the prediction correct.
}
\label{fig:reprompting}
\end{figure}

\faultless does catch and help the model correct genuine errors.
Figure~\ref{fig:reprompting} shows an example of how \faultless corrects GPT-5.5 about a missing function call.
However, GPT-5.5 often overreacts to \faultless' feedback by aggressively matching the decompiled code.
For instance, it may revert a struct field access operation back to pointer arithmetic, e.g. changing \cinline{root->left} to \cinline{*(long *)(a1 + 8)} in the second prediction.
In the most extreme instances, the new prediction is nearly identical to the decompiled code.
The especially high (90\%+) scores are partially a result of this kind of reward-hacking-like behavior.
The model may also change variable or function names back to the corresponding names in the decompiled code.
It is surprising that even a single reprompting instance can cause GPT-5.5 to completely disregard the detailed original prompt to produce code like the expected original.

Neural decompilation is ultimately a complex, multiobjective problem.
A neural decompiler has two main goals: (1) correctly representing the functionality in the binary or decompiled code, and (2) improving the code by re-introducing higher level abstractions, adding names, removing decompiler artifacts, and generally making the code easier to read and more similar to the original code.
\faultless offers significant improvement in the state of the art for checking (1), but does not make any attempt to evaluate (2).
Indeed, when a validation ``nonequivalent'' judgement is a result of a \faultless' incompleteness and the prediction is actually correct, reprompting can actually degrade the final prediction by undoing some of the improvements under the model's aggressive reward-hacking behavior.
This suggests that \faultless should be part of a larger suite of validation metrics which help push the model towards producing code that is both correct and higher quality.

Especially in combination with suitable improvement metrics, \faultless could be used as part of a reinforcement-learning setup to help improve neural decompiler predictions.

\begin{tcolorbox}[width=\columnwidth, boxsep=0pt, left=4pt, right=7pt, top=7pt, arc=4pt, boxrule=1pt, toprule=1pt, colback=white]%%
\noindent
Answer to \textbf{RQ7}: Re-prompting is very effective at improving \faultless validation scores and can help correct mispredictions, but models may overfit to the feedback and degrade the quality of the prediction.
\end{tcolorbox}

\section{Limitations}

Program equivalence is an extremely challenging task, and any program equivalence technique must navigate its inherent undecidability by making tradeoffs between soundness, completeness, and utility to end users.

\subsection{Soundness}
\label{sec:limitations:soundness}

While \faultless is designed with soundness in mind, in its current implementation, there are several well-defined areas in which it can report equivalence for nonequivalent functions.
None of these soundness limitations are fundamental; rather, they represent features that were deprioritized because the implementation effort required to support them was disproportionate with respect to their frequency in our datasets.

\subsubsection{Heap Base Case and Inductive Disassociation}

An important feature of \faultless' execution model is that $\phi$ symbolic variables are valid base addresses; writes relative to a $\phi$ variable represents a write on an arbitrary loop iteration.
However, because $\phi$ variables are distinct from the corresponding base-case arguments, a heapspace memory check from the base case will not incorporate the $\phi$ instruction's memory.
Therefore, \faultless reports the two functions in Figure~\ref{fig:base_case_phi_memory} equivalent, seeing no writes relative to the $n$ parameters.

In its simplest form, the solution is simple: when checking that memory accessible from two arguments is equivalent, the corresponding $\phi$ instructions' heaplets should be checked for equivalence as well. The heap already tracks the relationship between base cases and $\phi$ variables in the heap replay buffer (Section~\ref{sec:heap_replay_buffer}), making it easy to know which $\phi$ variables' heaplets to check. This can become more complex, especially when there are multiple $\phi$ instructions' heaplets to check.

Dissociation between the base cases' and $\phi$-variables' heaps is the most common soundness issue detected in Experiment 2 (Table~\ref{tab:soundness:faultless}) accounting for four of the five real soundness issues.

\begin{figure}
\begin{subfigure}[t]{0.50\textwidth}
\begin{lstlisting}[style=cstyle,basicstyle=\ttfamily\bfseries\scriptsize]
typedef struct node {
    int val;
    struct node * next;
} Node;
void initialize(Node *n) {
    while (n) {
        n->val = 1;
        n = n->next;
    }
}
\end{lstlisting}
\end{subfigure}
\begin{subfigure}[t]{0.49\textwidth}
\begin{lstlisting}[style=cstyle,basicstyle=\ttfamily\bfseries\scriptsize]
typedef struct node {
    int val;
    struct node * next;
} Node;
void zerolist(Node *n) {
    while (n) {
        n->val = 0;
        n = n->next;
    }
}
\end{lstlisting}
\end{subfigure}
\caption{Two differing writes to memory in a loop that \faultless does not detect when checking the equivalence of memory reachable from the input arguments.}
\label{fig:base_case_phi_memory}
\end{figure}

\begin{figure}
\begin{subfigure}[t]{0.50\textwidth}
\begin{lstlisting}[style=cstyle,basicstyle=\ttfamily\bfseries\scriptsize]
typedef struct node {
    int val;
    struct node * next;
} Node;
Node * newlist() {
    Node * tail = malloc(sizeof(Node));
    tail->val = 0;
    tail->next = NULL;
    Node * head = malloc(sizeof(Node));
    head->val = 0;
    head->next = tail;
    return head;
}
\end{lstlisting}
\end{subfigure}
\begin{subfigure}[t]{0.49\textwidth}
\begin{lstlisting}[style=cstyle,basicstyle=\ttfamily\bfseries\scriptsize]
typedef struct node {
    int val;
    struct node * next;
} Node;
Node * newlist() {
    Node * tail = malloc(sizeof(Node));
    tail->val = -1;
    tail->next = NULL;
    Node * head = malloc(sizeof(Node));
    head->val = 0;
    head->next = tail;
    return head;
}
\end{lstlisting}
\end{subfigure}
\caption{A pair of functions which are equivalent at one level of indirection but not at two levels of indirection; the tail node's \cinline{val} field is initialized with a \cinline{0} on the left and \cinline{-1} on the right. \faultless currently reports these as equivalent.}
\label{fig:recursive_heap_equivalence}
\end{figure}

\subsubsection{Recursive Heap Equivalence}

As part of the necessary equivalence conditions (Section~\ref{sec:defining_equivalence}), \faultless requires that memory reachable from the functions' arguments and callee arguments to be equivalent to the corresponding arguments in the other function.
However, at present \faultless does not apply this requirement recursively: it does not require that memory referenced at heap locations accessible from the input arguments are equivalent across the two functions.
Thus, \faultless will report the functions in Figure~\ref{fig:recursive_heap_equivalence} as equivalent.

The fix is involved but conceptually simple: for each addressable component of the memory locations read by the heaplet equivalence checker, the checker should be called recursively until a difference or an empty heaplet (no writes relative to that base address) is found.

This feature was deprioritized because programming paradigms around multi-indirection make it less of a pressing issue.
Many functions which recursively process linked data structures operate one ``layer'' (e.g. node) of a data structure at a time.
Or, alternatively, the whole data structure is processed sequentially in a loop, which results in special loop-$\phi$ related semantics.
No soundness issues of this type were discovered in the soundness testing experiment, though the \exebench dataset is light on \cinline{struct}s.

\begin{figure}
\begin{subfigure}[t]{0.50\textwidth}
\begin{lstlisting}[style=cstyle,basicstyle=\ttfamily\bfseries\scriptsize]
void divloop(int *x, int y) {
    while (*x > y) {
       *x = *x / 2;
    }
}
\end{lstlisting}
\caption{A function which mutates the same heap memory cell on each iteration of the loop.}
\label{fig:heap_induction:loop}
\end{subfigure}
\begin{subfigure}[t]{0.49\textwidth}
\begin{lstlisting}[style=cstyle,basicstyle=\ttfamily\bfseries\scriptsize]
void div(int *x, int y) {
    if (*x > y) {
        *x = *x / 2;
    }
}
\end{lstlisting}
\caption{A loop-free function which \faultless reports equivalent to Figure~\ref{fig:heap_induction:loop}}
\label{fig:heap_induction:if}
\end{subfigure}
\caption{A pair of examples which \faultless currently reports equivalent but which are not equivalent. Because the value controlling the iteration is on the heap rather than the stack in \cinline{divloop}, there is no $\phi$ instruction for \cinline{*x}.}
\label{fig:heap_induction}
\end{figure}

\subsubsection{Heap Induction}
\label{sec:heap_induction}

\faultless uses SSA $\phi$ instructions to identify variables which are mutated during the loop.
$\phi$ variables are used to help perform induction to prove loops equivalent.
This works for the vast majority of loops.

However, $\phi$ instructions apply to variables on the stack that are assigned to during the loop.
It is also possible to mutate the same \emph{heap} location on each iteration of a loop; Figure~\ref{fig:heap_induction:loop} shows an example.
This function has no $\phi$ instructions of any kind because all definitions (the two in the parameter lists) dominate all uses, so \faultless never performs induction.
Essentially, \faultless treats the loop as an if-statement as in Figure~\ref{fig:heap_induction:if}.

To handle this case, \faultless would have to perform induction on heap memory locations that are modified on multiple loop iterations.
Note that most in-loop heap modifications that occur in real code, like iterating over a linked list or array to initialize it, are not an issue because the modifications are defined in terms of $\phi$ instructions.
Instead, heap induction is required when there is a read from and write to the heap in a loop, where neither the base address nor offset is defined in terms of a $\phi$ symbolic variable: if this is the case, it means the memory location accessed is constant with respect to the loop and is thus acting like a stack variable which also has a constant address with respect to the loop.

A proposed, though unimplemented, solution to this issue is to place instructions of a new type called $\psi$ instructions at loop heads in which heap induction occurs.
These should have one argument: the address of the heap location that is modified during the loop.
To perform heap induction, for the base case, the values at the address must be equivalent when the loop iteration begins.
Then if they have the same value at the start of the loop, they should have the same values after the loop.
$\psi$ instructions should write mediloop and postloop symbolic $\psi$ variables to their memory addresses, just like $\phi$ instructions.
This is related to memory SSA, which we discuss further in Section~\ref{sec:indirect_memory_dependencies}.

In the soundness experiments, only one instance of heap induction was observed.
This is a rare programming pattern.

\subsection{Completeness}
\label{sec:limitations:completeness}

Because faultless is designed with soundness in mind, it is necessary incomplete.
Some completeness limitations are due to limitations in what \faultless can soundly assume in a context-independent environment.
Below are several important completeness limitations.
There are others besides those listed here; these are some of the most consequential.

\subsubsection{Indirect Dependencies through Memory}
\label{sec:indirect_memory_dependencies}

In the proof engine, the synchronization graph captures the dependencies between the proof obligations.
Dependencies reflect membership in the free-variable sets of synchronization nodes; a synchronization node is dependent on each other node which defines a symbolic variable in the former's free-variable set.
\faultless traverses a joint SSA and instruction-level control dependency graph to find these relationships.
However, not all dependencies are reflected in this graph.

\begin{figure}
\begin{subfigure}[t]{0.60\textwidth}
\begin{lstlisting}[style=cstyle,basicstyle=\ttfamily\bfseries\scriptsize]
struct node { char * s; struct node * next; };
void alloc_str(struct node * n, int len) {
    n->s = my_alloc(len);
    if (!n->s) {
        exit(0);
    }
}
\end{lstlisting}
\end{subfigure}
\begin{subfigure}[t]{0.39\textwidth}
\begin{lstlisting}[style=cstyle,basicstyle=\ttfamily\bfseries\scriptsize]

void func3(_QWORD *a1, int a2) {
    *a1 = func0(a2);
    if (!*a1) {
        exit(0);
    }
}
\end{lstlisting}
\end{subfigure}
\caption{A function and its hypothetical decompilation. They are equivalent, but \faultless will judge them nonequivalent due to the indirect heapspace dependency between \cinline{my_alloc} and \cinline{exit}.}
\label{fig:memory_dependency}
\end{figure}

The missing dependencies occur indirectly through the heap.
Figure~\ref{fig:memory_dependency} shows an example.
The function and its decompilation are judged as not equivalent because \faultless cannot prove the \cinline{exit} calls equivalent, despite the function in theory satisfying the necessary equivalence conditions.
The \cinline{exit} calls are control dependent on the \cinline{if} statement condition; to be considered equivalent two functions must be called under the same control flow conditions.
These \cinline{if} statements are in theory dataflow dependent on the values returned from \cinline{my_alloc} and \cinline{func0}.
But there is no SSA edge between the \cinline{if} condition and source of those calls.
Thus when \faultless goes to attempt to prove the \cinline{exit} functions equivalent, it compares their control flow conditions, by asking if  $(c_{\text{\cinline{my_alloc}}} = 0) = (c_{\text{\cinline{func0}}} = 0)$ is valid (always true), which it is not because \faultless does not know that $c_{\text{\cinline{my_alloc}}}$ actually represents the same thing as $c_{\text{\cinline{func0}}}$.

The best solution for this problem would be an analog of SSA for the heap, which would also solve the heap-induction soundness limitation (Section~\ref{sec:heap_induction}).
This is harder than regular SSA because def-use relations for heapspace memory is implicit and alias-dependent, but the problem is well studied in the literature~\cite{hssa,arrayssa,memoryssa} and could represent a promising future direction.

\subsubsection{Byte-Level Type Reinterpretation}

A key goal of \faultless is to see if the \emph{algorithm} captured by a neural decompiler's prediction---the partial sequence of operations it performs---is correct without focusing too much on the types in which that algorithm is expressed.
To this end, \faultless models primitive and pointer types---those which conceptually represent a single number from the perspective of the C language---as atomic, indivisible values.
Composite values, like \cinline{struct}s and arrays, are represented as aggregations of these atomic values at specific offsets.
This is advantageous when dealing with deterministic decompilers' necessarily imprecise type recovery and helping to provide compiler independence, bypassing trivial nonequivalence results due to type differences, and allowing for a seamless transition between math-int mode and bitvector mode.

However, this approach has its limits and also incurs some trade offs, particularly when processing decompiled code.
Decompiled code especially may abuse the memory layout of a given variable and represent multiple conceptual values in the original code as a single value.
For instance, consider the following snippet from a \realtype example: 

\noindent % Prevents paragraph indentation from pushing the layout out of bounds
\begin{minipage}[t]{0.60\linewidth}
\begin{lstlisting}[style=cstyle,basicstyle=\ttfamily\bfseries\scriptsize ]
char query[] = "aby";
int res = bs(query, p, 5);
\end{lstlisting}
\end{minipage}%
\hfill % Pushes the second minipage to the right edge
\begin{minipage}[t]{0.40\linewidth}%
\begin{lstlisting}[style=cstyle,basicstyle=\ttfamily\bfseries\scriptsize]
v2 = 7955041;
v3 = bs(&v2, v1, 5LL);
\end{lstlisting}
\end{minipage}
In the decompiled code, the four-byte (including the null terminator) string \cinline{"aby"} is packed into a four-byte integer variable.
\faultless sees the value in \cinline{query} as four distinct values while the value in \cinline{v2} as one value, so it rejects the calls to the binary search function as nonequivalent, despite the fact that they may be represented as identical byte sequences.

\subsubsection{Composite Globals in Validation}

Global variables are generally difficult for \faultless because there is no a priori mapping between global variables; rather, which globals in the left function correspond to those in the right must be inferred based on context.
In the validation task, where the right function is produced via deterministic decompilation, this problem is compounded when the global variable is a composite variable like a struct or an array because of the ways in which deterministic decompilers recover composite global variables.
In particular, different components (e.g. fields, elements) of the variable may be recovered as entirely different variables, sometimes just referencing a specific offset relative to the binary's base address, as occurs in the following original/decompiled snippets from a function in \realtype:

\noindent % Prevents paragraph indentation from pushing the layout out of bounds
\begin{minipage}[t]{0.60\linewidth}
\begin{lstlisting}[style=cstyle,basicstyle=\ttfamily\bfseries\scriptsize ]
 root->i_size = (2 * sizeof(struct dir_entry));
 root->i_time = time(((void *)0));
\end{lstlisting}
\end{minipage}%
\hfill % Pushes the second minipage to the right edge
\begin{minipage}[t]{0.40\linewidth}%
\begin{lstlisting}[style=cstyle,basicstyle=\ttfamily\bfseries\scriptsize]
qword_4044D0 = 774766593LL;
qword_4044D8 = 0LL;
\end{lstlisting}
\end{minipage}

\noindent
In the original C code, even if \faultless had the definition of the composite type, which \faultless does not assume for globals, the offsets to the fields are relative to the start of the global variable \cinline{root}, not the binary's base address as with the \cinline{qword_X} symbols in the decompiled code.
Additionally, \faultless sees the components as derived symbolic variables relative to a single unique variable base address, while the three \cinline{qword}s are seen as three separate global variables.
The decompiled code has no way to delineate where a composite global variable ends and another begins.

This is a context-independence and compiler-independence limitation; without knowing more about the global variables and the way they are laid out in memory, it's difficult to prove anything equivalent here.

\section{Conclusion}

Neural decompilation is a promising technique for helping overburdened reverse engineers to more quickly reconstruct abstractions in the target program.
Neural decompilation's greatest pitfall, however, is its probabilistic nature; predictions may be incorrect.
Assessing semantic correctness reduces to program equivalence.
\faultless is a novel program equivalence technique which is designed with neural decompilers in mind.
It is designed for two tasks, evaluation and validation, the former involving comparing a neurally predicted function with a gold-standard reference original version and the latter involving comparing the prediction with deterministic decompilation.
\faultless is largely context independent and compiler independent, requiring no information from the functions' calling contexts or dependencies aside from type definitions and it abstracts over most compiler-level details, modeling the C language at a higher level.
\faultless is substantially more capable than codealign~\cite{codealign}, an existing program equivalence technique for neural decompilation.
We show how \faultless is useful in evaluating a model's performance, comparing modeling approaches, ensuring that predictions are correct, and guiding the model to produce better predictions through re-prompting.

\bibliographystyle{ACM-Reference-Format}
\bibliography{references}

%\appendix

\end{document}